\documentclass[twocolumn,english,aps,prb]{revtex4}
\usepackage[T1]{fontenc}
\usepackage[latin9]{inputenc}
\usepackage{amsmath}
\usepackage{graphicx}
\usepackage{amssymb}

\makeatletter

\usepackage{amsbsy}

\usepackage{bm}

\makeatletter
\makeatletter
\makeatletter
\makeatletter
\makeatother
\makeatother
\makeatother 

\makeatother

\usepackage{babel}

\begin{document}

\title{Competing order and nature of the pairing state in the iron pnictides }

\author{Rafael M. Fernandes and J\"org Schmalian}

\affiliation{Ames Laboratory and Department of Physics and Astronomy, Iowa State
University, Ames, IA, 50010, USA}

\date{\today}

\begin{abstract}
We show that the competition between magnetism and superconductivity
can be used to determine the pairing state in the iron arsenides.
To this end we demonstrate that the itinerant antiferromagnetic phase
(AFM) and the unconventional $s^{+-}$ sign-changing superconducting
state (SC) are near the borderline of microscopic coexistence and
macroscopic phase separation, explaining the experimentally observed
competition of both ordered states. In contrast, conventional $s^{++}$
pairing is not able to coexist with magnetism. Expanding the microscopic
free energy of the system with competing orders around the multicritical point, we find that static
magnetism plays the role of an intrinsic interband Josephson coupling,
making the phase diagram sensitive to the symmetry of the Cooper pair
wavefunction. We relate this result to the quasiparticle excitation
spectrum and to the emergent SO$(5)$ symmetry of systems with particle-hole
symmetry. Our results rely on the assumption that the same electrons
that form the ordered moment contribute to the superconducting condensate and that the system is close to particle-hole symmetry.
We also compare the suppression of SC in different regions of the
FeAs phase diagram, showing that while in the underdoped side it is
due to the competition with AFM, in the overdoped side it is related
to the disappearance of pockets from the Fermi surface. 
\end{abstract}


\maketitle

\section{Introduction}

The recent discovery of iron arsenide superconductors\cite{Kamihara08,Chen08,Wang08,Rotter08}
brought renewed interest in the research of high-temperature superconductors.
With transition temperatures $T_{c}$ of more than $50$K in some
cases, these compounds present a very rich phase diagram displaying
superconducting (SC), antiferromagnetic (AFM) and structural order\cite{Luetkens09,Drew09,Ni09,Chu09}.
In some pnictides, such as $\mathrm{LaFeAs\left(O_{1-x}F_{x}\right)}$,
$\mathrm{PrFeAs\left(O_{1-x}F_{x}\right)}$, $\mathrm{\left(Sr_{1-x}Na_{x}\right)Fe_{2}As_{2}}$,
and $\mathrm{\left(Ba_{1-x}K_{x}\right)Fe_{2}As_{2}}$, the competing
SC and AFM phases seem to be separated by a first-order transition
and can only coexist in phase-separated macroscopic regions of the
sample\cite{Luetkens09,Rotundu09,Goko09,Fukazawa09,Hinkov09}. However,
in other compounds, like $\mathrm{Ba\left(Fe_{1-x}Co_{x}\right)_{2}As_{2}}$
and possibly\cite{Drew09} $\mathrm{SmFeAs\left(O_{1-x}F_{x}\right)}$,
local probes\cite{Laplace09,Julien09,Bernhard09} as well as bulk
measurements\cite{Ni09,Chu09,Ning09,Lester09,Pratt09,Christianson09}
demonstrated that SC and AFM coexist homogeneously. This coexistence,
however, is characterized by a competition of the two ordered states:
neutron diffraction experiments\cite{Pratt09,Christianson09} revealed
the dramatic suppression of the magnetization below $T_{c}$, to the
extent that reentrance of the non-magnetically ordered phase sets
in at low temperatures\cite{Fernandes10}.

Experiments have also demonstrated the itinerant character of the
magnetically ordered phase in the pnictides. In particular, optical
conductivity measurements show a considerable Drude weight as well
as a pronounced mid-infrared peak below the N\'eel transition temperature
$T_{N}$, consistent with the itinerant picture\cite{Hu08,Fernandes10_optics}. Furthermore,
band structure calculations reveal that the crystalline field is unable
to significantly split the energy levels in order to localize $3d$
electrons\cite{Singh08}. Also, several theoretical models demonstrate
the adequacy of the itinerant description\cite{Cvetkovic08,Korchunov08,Timm09,Eremin10}.
Therefore, in the iron arsenides, the same electrons that form the
superconducting condensate seem to be the ones that contribute to
the ordered moment.

\begin{figure}
\begin{centering}
\includegraphics[width=0.7\columnwidth]{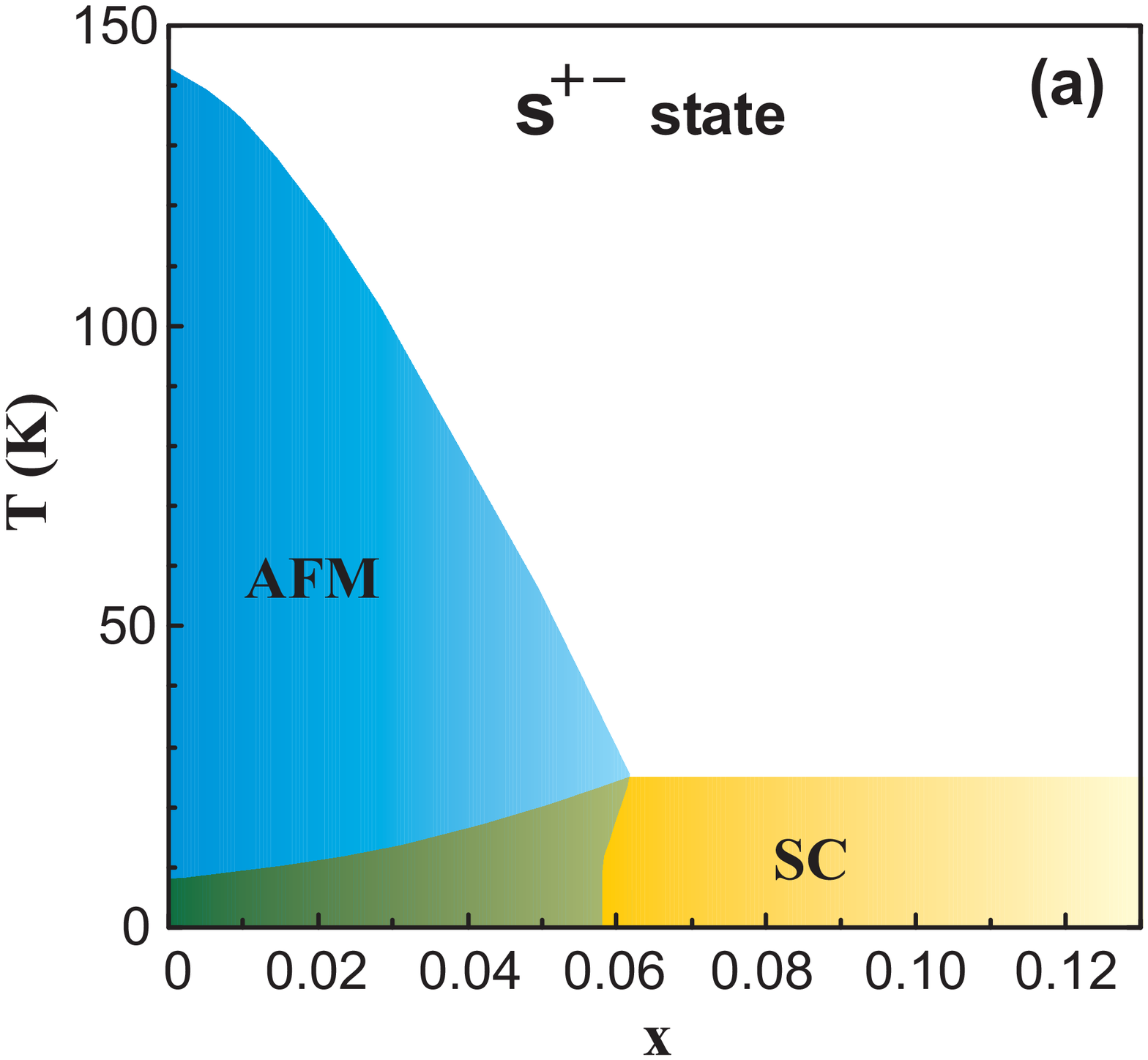} 
\par\end{centering}

\begin{centering}
\bigskip{}

\par\end{centering}

\begin{centering}
\includegraphics[width=0.7\columnwidth]{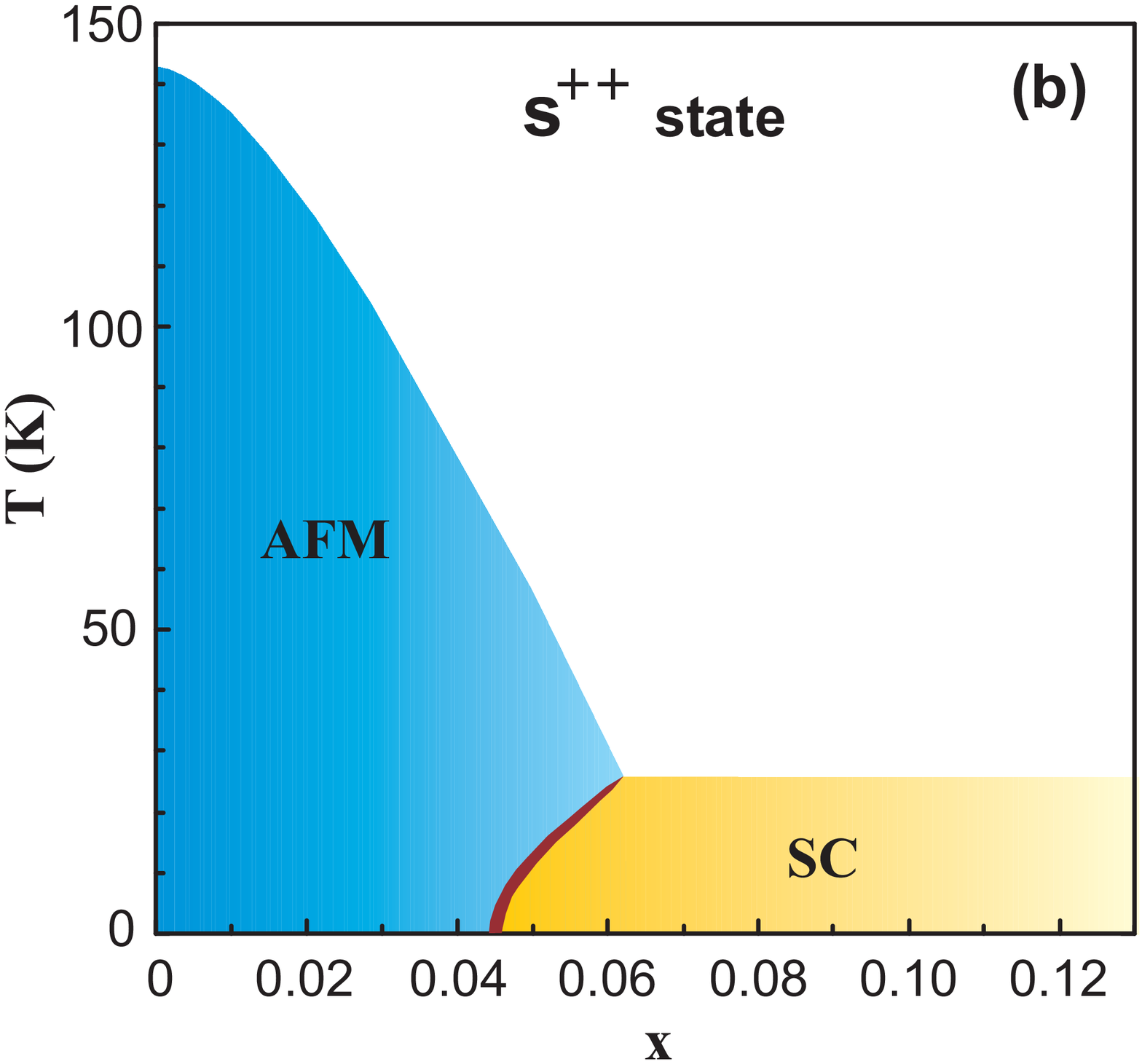} 
\par\end{centering}

\caption{Phase diagrams of $\mathrm{Ba\left(Fe_{1-x}Co_{x}\right)_{2}As_{2}}$
for a superconducting $s^{+-}$ state (a) and an $s^{++}$ state (b),
obtained by numerically solving the gap equations. The green region
denotes homogeneous, microscopic coexistence, whereas the dark red
region denotes heterogeneous, macroscopic coexistence. The band structure
parameters are discussed in Section IV-B. \label{fig_phase_diagrams} }

\end{figure}

The interplay between AFM and SC has been investigated in many contexts\cite{Baltensperger63,Bulaevskii80,Machida81,Grest81,Gabovich84,Gulacsi86,Machida88,Kulic95,Atkinson07,Das08},
including the pnictides\cite{Fernandes10,Vorontsov09,Tesanovic08,Parker_Vavilov09,Vavilov09,Vorontsov10,Ghaemi10}.
In this paper, following results from our previous work\cite{Fernandes10}
as well as from Refs.\cite{Vavilov09,Vorontsov10}, we investigate
in detail the connection between the competition of these two phases
and the pairing symmetry of the SC state. We demonstrate that coexistence
between SC and itinerant AFM in the pnictides for temperatures close to $T_N \simeq T_c$ is only possible if
the pairing state is unconventional, as proposed by models with purely
electronic pairing mechanisms\cite{Mazin08,Kuroki08,Chubukov08,Ikeda08,Wang09,Sknepnek09,Graser09,Junhua09,Mazin09}.
In particular, using a mean-field Hamiltonian for the competition
between AFM and SC and expanding the microscopic free energy in powers
of the order parameters, we show that a conventional $s^{++}$ SC
state does not allow a coexistence regime to be established around the point where the $T_N$ and $T_c$ lines meet, even for extreme values
of the band structure parameters. Meanwhile, the unconventional $s^{+-}$
state, whose gap function changes sign from one Fermi surface sheet
to the other, may or may not coexist with AFM, depending on the details
of the band structure dispersion relations. Specifically, for the
parameters of $\mathrm{Ba\left(Fe_{1-x}Co_{x}\right)_{2}As_{2}}$,
we find that the $s^{+-}$ state coexists with itinerant AFM, while
the $s^{++}$ state does not (see figure \ref{fig_phase_diagrams}). 

Our results rely solely on the general assumptions that the magnetism
is itinerant and that the band structure of the iron pnictides is not far from particle-hole symmetry, consisting
of two distinct sets of Fermi surface sheets\cite{Liu09}: hole pockets
located at the center of the Brillouin zone and electron pockets displaced
from the zone center by the magnetic ordering vector\textbf{ $\mathbf{Q}$}.
Additional details of the band structure, the dimensionality of the
system or the presence of intraband pairing interactions do not change
the conclusions. On the other hand, for a localized AFM state, the
free energy expansion reveals that coexistence is easily attained,
which is difficult to reconcile with the observation of phase separation
in some compounds. Our analysis also indicates that the onset of SC
has little effect on localized moments, which is at odds with experimental
observations as well, giving further evidence for the itinerant magnetism
in the pnictides. 

We also investigate in detail the origin of the strong dependence
of the phase diagram on the symmetry of the Cooper pair wavefunction.
Expressing the Ginzburg-Landau coefficients in terms of Feynman diagrams,
it becomes clear that the static staggered magnetic moment $\mathbf{m}$
plays the role of an intrinsic interband Josephson coupling. Specifically,
it corresponds to a term in the free energy of the form:

\begin{equation}
E_{J}\propto\mathbf{m}^{2}\left\vert \Delta_{1}\right\vert \left\vert \Delta_{2}\right\vert \cos\theta\label{Josephson}\end{equation}
where $\theta$ is the relative phase between the SC order parameters
of the two Fermi surface sheets, $\Delta_{1}$ and $\Delta_{2}$.
Thus, the coexistence state in some iron arsenides naturally carries
information about the relative phase of the Cooper pair wavefunctions,
which are usually accessible only through intricate and delicate interference
experiments\cite{Dale95,Chen10,Tsuei00}.

The quasiparticle excitation spectrum is substantially different for
distinct pairing symmetries. For the special case of particle-hole
symmetric bands, the system with competing magnetism and $s^{++}$
pairing have two distinct positive eigenvalues $E_{\mathbf{k}}$ with
$E_{\mathbf{k}}^{2}=\xi_{\mathbf{k}}^{2}+\left(M\pm\Delta\right)^{2}$,
whereas for $s^{+-}$ pairing the positive eigenvalues are degenerate:
$E_{\mathbf{k}}^{2}=\xi_{\mathbf{k}}^{2}+M^{2}+\Delta^{2}$. We find
that this special form of the excitation energy in the $s^{+-}$ case
implies that all quartic (and higher order) Ginzburg-Landau terms
must depend solely on the combination $\left(M^{2}+\Delta^{2}\right)$,
which is the root of the SO$(5)$ symmetry of the system and, ultimately,
what leads to the conclusion that AFM and SC are in the borderline
of coexistence and mutual exclusion. Inclusion of local moments add
new terms to the free energy, removing the system from this borderline
regime.

In this paper we also investigate some specific properties of the
coexistence AFM-SC phase. We find that the effects of Coulomb repulsion
on the magnetic superconducting phase are basically the same as in
the case of a pure multiband superconductor. Particularly, the $s^{+-}$
state is remarkably stable with respect to an uniform Coulomb repulsion.
We also studied analytically the shape of the reentrant N\'eel transition
line inside the SC state for low temperatures. At the mean-field level,
the finite SC gap introduces an overall energy scale that causes the
N\'eel line to have a divergent slope as $T\rightarrow0$. Quantum fluctuations,
which are relevant only in a very small region, end up suppressing
the reentrant behavior.

The competition between AFM and SC explains the suppression of $T_{c}$
in the underdoped side of the FeAs phase diagram. Here, we investigate
the suppression of SC on the overdoped side as well. We find that
the changes in the Fermi surface with doping are crucial to kill SC,
in agreement to ARPES measurements\cite{Sekiba09,Brouet09,Liu10Lif}.
In particular, the vanishing of one of the Fermi surface pockets marks
the onset of a regime where $T_{c}$ is strongly suppressed with doping.
In this regime, the $s^{+-}$ state is fragile and easily destroyed
by the Coulomb repulsion, contrasting to the situation where all Fermi
pockets are present.

The paper is organized as follows: in Section II we discuss the competition
between AFM and SC solely on phenomenological grounds. In Section
III we introduce the mean-field Hamiltonian and derive the gap equations,
the quasiparticle excitation spectrum and the free energy expansion.
Section IV is devoted to the application of the formalism developed
in Section III to the iron arsenides, and is divided in $5$ subsections.
In Section IV-A, we present the results in the special case of a particle-hole
symmetric band structure. Section IV-B contains both analytical and
numerical results for various band structures without particle-hole
symmetry. Phase diagrams for parameters describing $\mathrm{Ba\left(Fe_{1-x}Co_{x}\right)_{2}As_{2}}$
are presented. In Section IV-C, we briefly discuss the regime where
the sign of the coefficient that couples the AFM and SC order parameters
becomes negative, and how it can be avoided by the onset of incommensurate
AFM. Section IV-D discusses the effects of intraband interactions
and Coulomb repulsion. In Section IV-E, we determine analytically
the shape of the reentrant N\'eel transition line at low temperatures
and the corrections due to fluctuations. In Section V we solve the
same model presented in Section III, but now with localized magnetic
moments instead of itinerant AFM. Section VI discusses the suppression
of $T_{c}$ in the overdoped side of the pnictides phase diagram,
and how it is related to the doping evolution of the Fermi surface.
Section VII is devoted to the conclusions and, in Appendix A, we derive
the Ginzburg-Landau coefficients in terms of Feynman diagrams. Some
of the results have been published in a short publication\cite{Fernandes10}.

\section{Phenomenological analysis}

Regardless of the microscopic details, the competition between superconductivity
and antiferromagnetism near their finite temperature phase transitions
can be described in terms of a Ginzburg-Landau theory of coupled order
parameters. The homogeneous part of the free energy is given as

\begin{eqnarray}
F_{GL}\left(\Delta,\mathbf{M}\right) & = & \int d^{d}r\left(\frac{a_{s}}{2}\left\vert \Delta\right\vert ^{2}+\frac{u_{s}}{4}\left\vert \Delta\right\vert ^{4}+\frac{\gamma}{2}\left\vert \Delta\right\vert ^{2}\mathbf{M}^{2}\right.\nonumber \\
 &  & \left.+\frac{a_{m}}{2}\mathbf{M}^{2}+\frac{u_{m}}{4}\mathbf{M}^{4}\right),\label{GL}\end{eqnarray}
 where $\Delta$ and $\mathbf{M}$ denote the SC and AFM order parameters,
respectively. As usual, $\Delta$ is a complex order parameter, characterized
by an amplitude and a phase, and $\mathbf{M}$ is a three component
vector. The leading term in the order parameter competition is characterized
by the coefficient $\gamma>0$, where the sign of $\gamma$ reflects
that both ordered states compete. As usual, the quadratic coefficients
are given by $a_{m}=a_{m,0}\left(T-T_{N,0}\right)$ and $a_{s}=a_{s,0}\left(T-T_{c,0}\right)$
and change sign at $T_{N,0}$ and $T_{c,0}$, denoting the N\'eel and
SC transition temperatures without order parameter competition. We
consider the situation where the transitions for $\gamma=0$ are second
order, i.e. the quartic coefficients $u_{m}$ and $u_{s}$ are positive.

Furthermore, we consider that $T_{N,0}\left(x\right)$ and $T_{c,0}\left(x\right)$
vary as function of a physical parameter $x$ that could be pressure,
electron density or magnetic field. In case where both transitions
meet at $x=x^{\ast}$, i.e. for \begin{equation}
T^{\ast}\equiv T_{N,0}\left(x^{\ast}\right)=T_{c,0}\left(x^{\ast}\right),\end{equation}
we have a multicritical point $\left(x^{\ast},T^{\ast}\right)$ in
the phase diagram. The vicinity of this multicritical point is the
regime where a simultaneous expansion of the order parameters is allowed.
The mean-field analysis of Eq.\ref{GL} allows for two options for
the phase diagram near $\left(x^{\ast},T^{\ast}\right)$, depending
if $\gamma^{2}>u_{m}u_{s}$ or $\gamma^{2}<u_{m}u_{s}$. Since we
are interested in $\gamma>0$, it is convenient to define the dimensionless
quantity\cite{comment01} \begin{equation}
g\equiv\frac{\gamma}{\sqrt{u_{m}u_{s}}}-1.\label{g_parameter}\end{equation}

\begin{figure*}
\begin{centering}
\includegraphics[width=1.85\columnwidth]{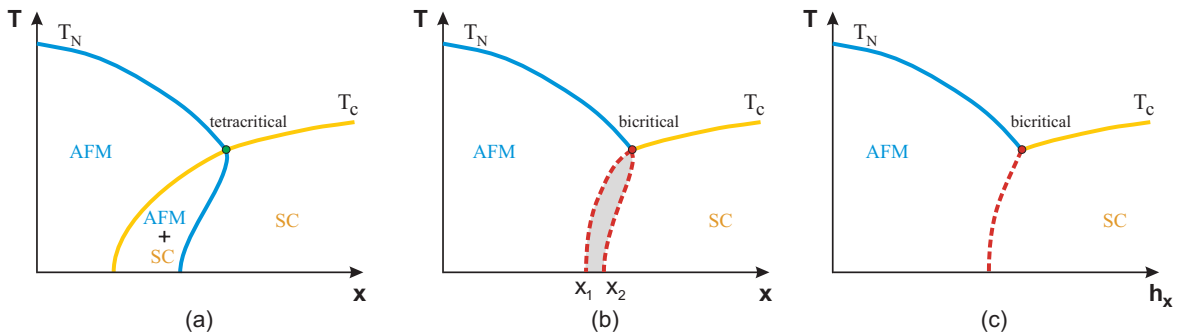}
\par\end{centering}

\caption{Schematic phase diagrams $\left(x,T\right)$ for competing AFM and
SC orders. Here, $x$ is a generic physical parameter and $T$ is
the temperature. Solid (dashed) lines denote second-order (first-order)
phase transitions. For $g<0$, there is a tetracritical point and
a region of homogeneous coexistence (a), whereas for $g>0$ there
is a bicritical point (b and c). If $x$ changes discontinuously across
the first order transition (panel b), forming a region of heterogeneous
coexistence for $x_{1}<x<x_{2}$ (shaded area), then its conjugate
variable $h_{x}$ changes continuously and the phase diagram has only
one first-order line (panel c). \label{fig_tetra_bicritical}}

\end{figure*}

Thus, the nature of the phase diagram is determined solely by the
quartic coefficients in the Ginzburg-Landau expansion Eq.\ref{GL}.
For $g<0$ (i.e. $0<\gamma<\sqrt{u_{s}u_{m}}$), $\left(x^{\ast},T^{\ast}\right)$
is a tetracritical point where two second order phase lines cross,
leading to a regime in the phase diagram where simultaneous AFM and
SC order occurs homogeneously within the sample, see Fig.\ref{fig_tetra_bicritical}a. In this
regime both phases compete, but do not exclude each other. On the
other hand, if $g>0$ (i.e $\gamma>\sqrt{u_{s}u_{m}}$) the phase
competition is sufficiently strong that both phases are separated
by a first order transition that terminates at the bicritical point
$\left(x^{\ast},T^{\ast}\right)$. Notice that if the parameter $x$
jumps discontinuously from $x_{1}$ to $x_{2}$ at the first order
transition, there is an intermediate regime $x_{1}<x<x_{2}$ of heterogeneous
phase coexistence, see Fig.\ref{fig_tetra_bicritical}b. A sharp line of first order transitions
occurs if one considers the phase diagram as function of $h_{x}$,
the variable that is thermodynamically conjugate to $x$, see Fig.\ref{fig_tetra_bicritical}c. 
Critical fluctuations, that go beyond this mean-field analysis,
change the universal exponents near the critical temperatures and
the slopes of the phase lines near $\left(x^{\ast},T^{\ast}\right)$.
However, neither the generic behavior shown in Fig.1 nor the quantitative
criterion based on the sign of $g$ are changed by fluctuations\cite{Kosterlitz76,Aharony03}.

If we consider $g<0$, both order parameters can be finite simultaneously.
This regime is often referred to as \emph{coexistence} of AFM and
SC, referring to coexistence of order. This should not be confused
with phase coexistence in the thermodynamic sense. The area
in Fig.\ref{fig_tetra_bicritical}a below the tetracritical point is a single thermodynamic
phase characterized by two order parameters that are simultaneously
finite. Similarly, the tetracritical point is not a point where four
phases coexist (which would not be allowed by Gibbs phase rule), but
a point where the system is in a single phase and both order parameters
are infinitesimal simultaneously. Below the bicritical point, coexistence
of thermodynamic phases only occurs for $x_{1}<x<x_{2}$, where macroscopic
AFM and SC regions occur together in the sample. We use the term \emph{homogeneous
coexistence} of AFM and SC order below the tetracritical point to
refer to coexisting order and \emph{heterogeneous coexistence} below
the bicritical point to refer to coexistence of phases.

From the Ginzburg-Landau expression (\ref{GL}), we obtain the temperature
dependence of the magnetic moment in the SC phase in the case of homogeneous
coexistence:

\begin{equation}
\mathbf{M}^{2}\left(T\right)=\frac{a_{m,0}u_{s}\left(T_{N,0}-T\right)+a_{s,0}\gamma\left(T-T_{c,0}\right)}{\sqrt{u_{m}u_{s}}-\gamma}\label{magnetic_moment_SC}\end{equation}


%


Without phase competition both order parameters decrease as function
of temperature, $d\mathbf{M}^{2}/dT<0$ and $d\Delta/dT<0$. Phase
competition can change this behavior. For instance, if

\begin{equation}
a_{s,0}\gamma>a_{m,0}u_{s},\label{condition_suppression}\end{equation}
it follows from Eq. (\ref{magnetic_moment_SC}) that $d\mathbf{M}^{2}/dT>0$
once superconductivity sets in. Thus, below $T_{c}$, the ordered
moment decreases with decreasing temperature, as was observed in neutron
diffraction experiments\cite{Pratt09,Christianson09} in both $\mathrm{Ba\left(Fe_{1-x}Co_{x}\right)_{2}As_{2}}$
and\cite{Kreyssig10} $\mathrm{Ba\left(Fe_{1-x}Rh_{x}\right)_{2}As_{2}}$.
It is interesting that the same condition implies a back-bending of
the antiferromagnetic phase boundary upon entering the superconducting
state, see Fig. 1a. To demonstrate this we write for the bare N\'eel
temperature $T_{N,0}(x)=T_{c,0}\left(1+f\left(x\right)\right)$, with
$df/dx<0$ and $f(x^{\ast})=0$. Thus, $T_{N,0}$ is a monotonic decreasing
function of $x$ and meets the SC phase line at the carrier density
or pressure value $x^{\ast}$. Without restriction we assume that
$T_{c,0}$ is $x$-independent near $x^{\ast}$. From the Ginzburg-Landau
expansion Eq.(\ref{GL}) follows that the N\'eel temperature $T_{N}$
inside the SC phase is given by:

\begin{equation}
T_{N}=T_{c,0}\left(1-f\left(x\right)\frac{a_{m,0}u_{s}}{a_{s,0}\gamma-a_{m,0}u_{s}}\right).\label{TN_inside_SC}\end{equation}

Since $df/dx<0$, it follows that $dT_{N}/dx>0$ in case Eq.(\ref{condition_suppression})
is valid. Thus close to $x^{\ast}$ one finds reentrance of the paramagnetic
phase below the SC transition temperature. In $\mathrm{Ba\left(Fe_{1-x}Co_{x}\right)_{2}As_{2}}$
this was indeed observed\cite{Pratt09,Christianson09}. The observation\cite{Kreyssig10}
of $d\mathbf{M}^{2}/dT>0$ in $\mathrm{Ba\left(Fe_{1-x}Rh_{x}\right)_{2}As_{2}}$
implies that the phase line for this material must bend back as well.
Furthermore, the SC transition temperature inside the AFM phase is
given by:

\begin{equation}
T_{c}=T_{c,0}\left(1+f\left(x\right)\frac{a_{m,0}\gamma}{a_{m,0}\gamma-a_{s,0}u_{m}}\right).\label{Tc_inside_AFM}\end{equation}

Using the condition Eq.(\ref{condition_suppression}) and the fact
that $\gamma<\sqrt{u_{m}u_{s}}$ ($g<0$) it automatically holds that
$a_{m,0}\gamma<a_{s,0}u_{m}$, implying $dT_{c}/dx>0$. Hence, the
SC transition temperature is suppressed inside the AFM phase.

A very interesting limit is $g=0$, i.e. at the transition between
the tetracritical point and bicritical point. Focusing on this multicritical
point, where $a_{s}$ and $a_{m}$ change sign simultaneously, one
can introduce the five-component vector\begin{equation}
\overrightarrow{\mathbf{N}}\mathbf{=}\left(\frac{a_{s}}{a_{m}}\right)^{1/4}\left(\ \mathrm{Re}\Delta,\ \mathrm{Im}\Delta,\frac{a_{s}}{a_{m}}\mathbf{M}\right).\end{equation}

In case where the additional condition $u_{s}=a_{s}^{2}u_{m}/a_{m}^{2}$
is fulfilled it follows that the free-energy can be written as: \begin{equation}
F_{GL}\left(\Delta,\mathbf{M}\right)=\int d^{d}r\left(\frac{a}{2}\overrightarrow{\mathbf{N}}^{2}+\frac{u}{4}\overrightarrow{\mathbf{N}}^{4}\right)\end{equation}
where $a=\mathrm{sign}\left(a_{s}\right)\sqrt{\left\vert a_{s}\right\vert \left\vert a_{m}\right\vert }$
and $u=\sqrt{u_{s}u_{m}}$. This is the SO$\left(5\right)$ symmetric
form of the Ginzburg-Landau energy that was first proposed by Zhang
to describe the physics of the cuprate superconductors\cite{Zhang97,Demler04}. In the context
of the pnictides, it was shown\cite{Podolski09} that a model Hamiltonian
similar to the one used in this paper is invariant with respect to
a global SO$\left(6\right)$ symmetry that contains, in addition to
the AFM and SC order parameters, an imaginary density wave state.
Below we will see that there is evidence that the the pnictides are
indeed strongly affected by such an enhanced symmetry.

\section{Microscopic model}

\noindent So far we have analyzed the problem of coexistence between
SC and AFM only on phenomenological grounds, which gave us interesting
and general information about the phase diagram. Next we develop a
microscopic model that captures the essential aspects of the iron
arsenides to determine their detailed phase diagram. We will also
make explicit contact to the Ginzburg-Landau theory and determine
the coefficients of the order parameter expansion to obtain the behavior
close to the transition temperatures.

We start from the Hamiltonian: \begin{equation}
\mathcal{H}=\mathcal{H}_{0}+\mathcal{H}_{\mathrm{AFM}}+\mathcal{H}_{\mathrm{SC}}.\label{full_Hamiltonian}\end{equation}

The non-interacting part $\mathcal{H}_{0}$ describes two bands shifted
by the momentum $\mathbf{Q}$ relative to each other:

\begin{equation}
\mathcal{H}_{0}=\sum_{\mathbf{k}\sigma}\left(\varepsilon_{1,\mathbf{k}}-\mu\right)c_{\mathbf{k}\sigma}^{\dagger}c_{\mathbf{k}\sigma}+\sum_{\mathbf{k}\sigma}\left(\varepsilon_{2,\mathbf{\mathbf{k}+Q}}-\mu\right)d_{\mathbf{k+Q}\sigma}^{\dagger}d_{\mathbf{k+Q}\sigma}.\label{H0}\end{equation}

\begin{figure}
\begin{centering}
\includegraphics[width=0.75\columnwidth]{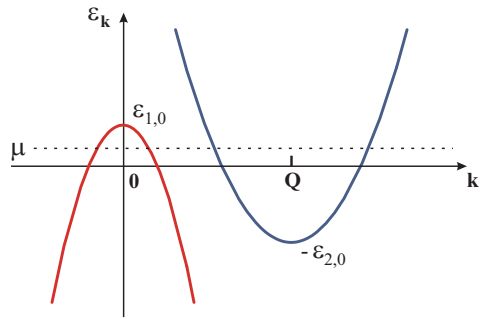}
\par\end{centering}

\caption{Schematic representation of the two-band structure used here. It consists
of a circular hole band (band 1, red line) in the center of the Brillouin
zone and an elliptical electron band (band 2, blue line) shifted by
the ordering vector $\mathbf{Q}$ from the hole band. The chemical
potential $\mu$ is positive for electron doping. \label{fig_band_structure}}

\end{figure}

We consider only one hole band located in the center of the Brillouin
zone with dispersion $\varepsilon_{1,\mathbf{k}}$, and one electron
band, shifted by $\mathbf{Q}$ from the hole band, with dispersion
$\varepsilon_{2,\mathbf{k}}$. To keep the discussion as simple as
possible, but still capturing the basic properties of these materials,
we consider a circular hole band and an elliptical electron band:

\begin{eqnarray}
\varepsilon_{1,\mathbf{k}} & = & \varepsilon_{1,0}-\frac{k^{2}}{2m},\nonumber \\
\varepsilon_{2,\mathbf{k}+\mathbf{Q}} & = & -\varepsilon_{2,0}+\frac{k_{x}^{2}}{2m_{x}}+\frac{k_{y}^{2}}{2m_{y}},\label{band_dispersion}\end{eqnarray}
where $\varepsilon_{\alpha,0}$ is the energy offset (see figure \ref{fig_band_structure}).
Such a choice is motivated by angle-resolved photoemission spectroscopy\cite{Liu09}
(ARPES) as well as by tight-binding fittings to first-principle band
structure calculations. The operator $c_{\mathbf{k}\sigma}^{\dagger}$
($d_{\mathbf{k+Q}\sigma}^{\dagger}$) creates an electron with momentum
$\mathbf{k}$ ($\mathbf{k+Q}$) and spin $\sigma$ in the hole (electron)
band. The chemical potential is denoted by $\mu$, and we define $\xi_{\alpha,\mathbf{k}}=\varepsilon_{\alpha,\mathbf{k}}-\mu$.
Frequently, it will be possible to gain explicit analytic insight
by considering the limit of particle-hole symmetry, where $\varepsilon_{0}\equiv\varepsilon_{1,0}=\varepsilon_{2,0}$,
$m_{x}=m_{y}=m$ and $\mu=0$. In this case the hole and electron
Fermi surfaces are identical, leading to perfect nesting. In what
follows we supplement our numerical analysis of the problem with band
structure Eq.(\ref{band_dispersion}) by analytical results at or
near the limit of particle-hole symmetry. Even though all five iron
$d$-orbitals contribute to the states at the Fermi surface of the
iron arsenides, the physics of the competition between the antiferromagnetic
and the superconducting states is well captured by this effective
two-band model\cite{Fernandes10}.

The same electrons that form the Fermi surface are assumed to be responsible
for the magnetism of the system through an electronic interband Coulomb
interaction $I$, leading to excitonic itinerant antiferromagnetism\cite{Keldysh65,Cloizeaux65,Timm09}:

\begin{equation}
\mathcal{H}_{\mathrm{AFM}}=I\sum_{\mathbf{k,k}^{\prime},\mathbf{q}}\sum_{s,s^{\prime}}c_{\mathbf{k}s}^{\dagger}\boldsymbol{\sigma}_{ss^{\prime}}d_{\mathbf{k+q}s^{\prime}}\cdot d_{\mathbf{k}^{\prime}s}^{\dagger}\boldsymbol{\sigma}_{ss^{\prime}}c_{\mathbf{k}^{\prime}-\mathbf{q}s^{\prime}}.\label{H_AFM_full}\end{equation}

Here $\sigma_{ss^{\prime}}^{(i)}$ denotes the $\left(ss^{\prime}\right)$
element of the $i$-th Pauli matrix, with $s=\pm1$. In the weak-coupling
limit, we perform a Hartree-Fock decoupling that leads, except for
a constant shift in energy, to the effective single particle Hamiltonian

\begin{equation}
\mathcal{H}_{\mathrm{AFM}}=-\frac{1}{N} \sum_{\mathbf{\mathbf{k}}s}sM\left(c_{\mathbf{k}s}^{\dagger}d_{\mathbf{k+Q}s}+d_{\mathbf{k+Q}s}^{\dagger}c_{\mathbf{k}s}\right),\label{H_AFM}\end{equation}
where $N$ is the system size. $M$ denotes the antiferromagnetic
gap opened at momenta $\mathbf{k}_{0}$ that are Bragg scattered due
to magnetic order, $\xi_{1,\mathbf{k}_{0}}=\xi_{2,\mathbf{k}_{0}\mathbf{+Q}}$.
In general, for small $M$, the Fermi surface is only partially gapped 
and the magnetic state is metallic. For large enough $M$, however, the Fermi surface can become completely gapped. 
In the case of perfect nested bands, an infinitesimal antiferromagnetic gap is
already able to gap the entire Fermi surface. Note that $M=\left|\mathbf{M}\right|$ is proportional to
the amplitude of the staggered magnetization $\mathbf{m}$ and given by:

\begin{equation}
M=\frac{I}{2N}\sum_{\mathbf{k},s}s\left\langle c_{\mathbf{k}s}^{\dagger}d_{\mathbf{k+Q}s}\right\rangle =I\: m.\label{AFM_gap}\end{equation}

Besides the magnetic interaction $I$, the electrons are also subject
to a pairing interaction $V_{\alpha\beta}$, where $\alpha,\beta=1,2$
are band indices. In case of pure interband interaction $V_{\alpha\beta}=V\left(1-\delta_{\alpha\beta}\right)$
the Hamiltonian becomes

\begin{equation}
\mathcal{H}_{\mathrm{SC}}=V\sum_{\mathbf{k,k}^{\prime},\mathbf{q}}c_{\mathbf{k+q}\mathbf{\uparrow}}^{\dagger}c_{-\mathbf{k\downarrow}}^{\dagger}d_{-\mathbf{k}^{\prime}-\mathbf{q\uparrow}}d_{\mathbf{k}^{\prime}\uparrow}.\label{H_SC_full}\end{equation}

Below we demonstrate that the introduction of an intraband pairing
interaction does not change the main conclusions of our paper. $\Delta_{\alpha}$
is the superconducting gap of band $\alpha$ which, given the interband
coupling $V$, is due to the action of the electrons in the opposite
band $\bar{\alpha}$. Thus, $\Delta_{1}$ and $\Delta_{2}$ are determined
by the two gap-equations:

\begin{eqnarray}
\Delta_{2} & = & -\frac{V}{N}\sum_{\mathbf{k}}\left\langle c_{\mathbf{k}\uparrow}^{\dagger}c_{-\mathbf{k}\downarrow}^{\dagger}\right\rangle ,\nonumber \\
\Delta_{1} & = & -\frac{V}{N}\sum_{\mathbf{k}}\left\langle d_{\mathbf{k+Q}\uparrow}^{\dagger}d_{-\mathbf{k}-\mathbf{Q}\downarrow}^{\dagger}\right\rangle ,\label{SC_gap}\end{eqnarray}

The expectation values are determined with the mean field Hamiltonian:\begin{eqnarray}
\mathcal{H}_{\mathrm{SC}} & = & -\sum_{\mathbf{k}}\left(\Delta_{1}c_{\mathbf{k}\uparrow}^{\dagger}c_{-\mathbf{k}\downarrow}^{\dagger}+h.c.\right)\label{H_SC}\\
 &  & -\sum_{\mathbf{k}}\left(\Delta_{2}d_{\mathbf{k+Q}\uparrow}^{\dagger}d_{-\mathbf{k}-\mathbf{Q}\downarrow}^{\dagger}+h.c.\right).\nonumber \end{eqnarray}

The mean-field Hamiltonian formed by the sum of Eqs. (\ref{H0}),
(\ref{H_AFM}) and (\ref{H_SC}) is quadratic and can be diagonalized
analytically, yielding the self-consistent gap equations:

\begin{eqnarray}
\Delta_{\alpha} & = & -\frac{V}{N}\sum_{\mathbf{k},a}K_{\mathbf{k},\alpha}^{\Delta}\tanh\left(\frac{\beta E_{a,\mathbf{k}}}{2}\right)\nonumber \\
M & = & \frac{I}{N}\sum_{\mathbf{k},a}K_{\mathbf{k},\alpha}^{M}\tanh\left(\frac{\beta E_{a,\mathbf{k}}}{2}\right)\label{gap_equations}\end{eqnarray}
with kernels: \begin{eqnarray}
K_{\mathbf{k},\alpha}^{\Delta} & = & \frac{\Delta_{\bar{\alpha}}\left(E_{a,\mathbf{k}}^{2}-\Delta_{\alpha}^{2}-\xi_{\alpha,\mathbf{k}}^{2}\right)+M^{2}\Delta_{\alpha}}{2E_{a,\mathbf{k}}\left(E_{a,\mathbf{k}}^{2}-E_{\bar{a},\mathbf{k}}^{2}\right)}\nonumber \\
K_{\mathbf{k},\alpha}^{M} & = & \frac{M\left(E_{a,\mathbf{k}}^{2}+\Delta_{1}\Delta_{2}+\xi_{1,\mathbf{k}}\xi_{2,\mathbf{k}}-M^{2}\right)}{2E_{a,\mathbf{k}}\left(E_{a,\mathbf{k}}^{2}-E_{\bar{a},\mathbf{k}}^{2}\right)}\end{eqnarray}

The excitation energies $E_{a,\mathbf{k}}$ ($a=1,2$) are the positive
eigenvalues of a state with simultaneous magnetic and superconducting
order:

\begin{equation}
E_{a,\mathbf{k}}^{2}=\frac{1}{2}\left(\Gamma_{\mathbf{k}}\pm\sqrt{\Gamma_{\mathbf{k}}^{2}+\Omega_{\mathbf{k}}+\delta_{\mathbf{k}}}\right)\label{excitation_energies}\end{equation}
with $\Gamma_{\mathbf{k}}=2M^{2}+\Delta_{1}^{2}+\Delta_{2}^{2}+\varepsilon_{\mathbf{k,}1}^{2}+\varepsilon_{\mathbf{k+Q,}2}^{2}$
and $\Omega_{\mathbf{k}}=-4\left(\varepsilon_{\mathbf{k,}1}^{2}+\Delta_{1}^{2}\right)\left(\varepsilon_{\mathbf{k+Q,}2}^{2}+\Delta_{2}^{2}\right)$
as well as $\delta_{\mathbf{k}}=8M^{2}\left(\Delta_{1}\Delta_{2}+\varepsilon_{\mathbf{k,}1}\varepsilon_{\mathbf{k+Q,}2}-M^{2}/2\right)$.
The free energy density of a system with SC and AFM long range order
that results from this analysis is: \begin{eqnarray}
f\left(M,\Delta_{\alpha}\right) & = & \frac{2}{I}M^{2}-\frac{1}{V}\left(\Delta_{1}^{\ast}\Delta_{2}+\Delta_{2}^{\ast}\Delta_{1}\right)\nonumber \\
 &  & -\frac{2T}{N}\sum_{\mathbf{k,}a}\log\left(2\cosh\left(\frac{E_{a,\mathbf{k}}}{2k_{B}T}\right)\right).\label{F}\end{eqnarray}

The superconducting order parameters $\ \Delta_{1}\ $and $\ \Delta_{2}$
of the two bands and the staggered moment $\propto M$ are obtained
by minimizing $f\left(M,\Delta_{\alpha}\right)$. The gap equations
(\ref{gap_equations}) follow as the stationary points $\partial f\left(M,\Delta_{\alpha}\right)/\partial\Delta_{\alpha}=\partial f\left(M,\Delta_{\alpha}\right)/\partial M=0$.

Before we analyze the impact of magnetic long range order on the pairing
state we discuss the gap equations (\ref{gap_equations}) in the limit
$M=0$. Here, we perform the momentum integration by introducing the
density of states $\rho_{1}$ and $\rho_{2}$ of the two bands. The
condition for $T_{c}$ is that the largest eigenvalue of \begin{equation}
\Lambda=\left(\begin{array}{cc}
0 & -V\rho_{2}\\
-V\rho_{1} & 0\end{array}\right).\label{pairing matrix}\end{equation}
is positive and equals to $1/\ln\left(W/\left(\alpha T_{c}\right)\right)$,
were $\alpha=\pi e^{-\gamma_{E}}/2$ , $\gamma_{E}$ is Euler's constant
and $W$ is an upper energy cutoff for the pairing interaction. Clearly,
the eigenvalues of $\Lambda$ are $\lambda_{\pm}=\pm V\sqrt{\rho_{1}\rho_{2}}$.
The pairing state is determined by the corresponding eigenvector \begin{equation}
\left(\Delta_{1},\Delta_{2}\right)\propto\frac{1}{\sqrt{\rho_{1}+\rho_{2}}}\left(\sqrt{\rho_{2}},\mp\sqrt{\rho_{1}}\right)\end{equation}

Thus, for $V<0$, $\lambda_{-}$ is the largest eigenvalue and the
gap equation has a solution where $\Delta_{1}$ and $\Delta_{2}$
on the two sheets of the Fermi surface have the same sign. This is
called the $s^{++}$ state and is analogous to the pairing state of
the multiband superconductor\cite{Canfield03} $\mathrm{MgB_{2}}$.
It is the natural state that arises as a result of conventional electron-phonon
interactions. On the other hand, the gap equations also allow for
a solution for $V>0$, with $\Delta_{1}$ and $\Delta_{2}$ having
different signs on distinct Fermi surface sheets. This $s^{+-}$-state
results from purely electronic interactions,\cite{Mazin08,Kuroki08,Chubukov08,Cvetkovic08,Wang09,Sknepnek09,Graser09,Junhua09,Mazin09} 
i.e. it is the natural analog to the $d_{x^{2}-y^{2}}$-pairing state in the cuprates with
a single Fermi surface sheet.
As the relative sign of $\Delta_{1}$ and $\Delta_{2}$ is $-1$,
this is an unconventional SC state, even though it is in the same
irreducible representation $A_{1g}$ of the symmetry group $D_{4h}$
as the $s^{++}$-state.

In case where $\Delta_{1}$ and $\Delta_{2}$ have the same sign ($s^{++}$-state),
the excitation energies $\pm E_{2,\mathbf{k}}$ possibly have nodes\cite{Parker_Vavilov09}.
The nodes are located at the set of points $\mathbf{k}_{n}$ that
satisfy simultaneously the equations:

\begin{eqnarray}
\xi_{1,\mathbf{k}_{n}} & = & \pm\sqrt{\frac{\Delta_{1}}{\Delta_{2}}}\left(M^{2}-\Delta_{1}\Delta_{2}\right)^{1/2}\nonumber \\
\xi_{2,\mathbf{k}_{n}\mathbf{+Q}} & = & \pm\sqrt{\frac{\Delta_{2}}{\Delta_{1}}}\left(M^{2}-\Delta_{1}\Delta_{2}\right)^{1/2}\label{cosnd}\end{eqnarray}
determined from the condition $E_{2,\mathbf{k}}=0$. Obviously, one
condition for nodes to exist is $M^{2}\geq\Delta_{1}\Delta_{2}$.
For $\Delta_{1}=\Delta_{2}$ the condition for nodes in the antiferromagnetic
state corresponds to $\xi_{1,\mathbf{k}_{n}}=\xi_{2,\mathbf{k}_{n}\mathbf{+Q}}$
i.e. where Bragg scattering due to antiferromagnetism is large. However,
nodes are not guaranteed to emerge. For example, in the case of particle
hole symmetry (implying perfect nesting of the Fermi surface)\begin{equation}
\xi_{\mathbf{k}}\equiv\xi_{1,\mathbf{k}}=-\xi_{2,\mathbf{k+Q}}\end{equation}
it holds for $I>\left\vert V\right\vert $ that $\Delta\equiv\Delta_{1}=\pm\Delta_{2}=\frac{\left\vert V\right\vert }{I}M<M$
and the above equations cannot be fulfilled simultaneously. This also
follows if one explicitly considers the eigenvalues for $s^{++}$
pairing in the limit of particle-hole symmetry:\begin{equation}
E_{a,\mathbf{k}}^{2}=\xi_{\mathbf{k}}^{2}+\left(M\pm\Delta\right)^{2}\end{equation}
which are fully gapped for $M\neq\Delta$. Only for $\left\vert V\right\vert =I$
follows an entirely gapless eigenvalue $E_{2,\mathbf{k}}$, consistent
with the condition Eq.(\ref{cosnd}). In distinction, the eigenvalues
for $s^{+-}$ pairing and particle-hole symmetry are fully gapped
and doubly degenerate: \begin{equation}
E_{a,\mathbf{k}}^{2}=\xi_{\mathbf{k}}^{2}+M^{2}+\Delta^{2}.\end{equation}

It is also interesting to note that, if one considers the simplification $\Delta=\Delta_1=-\Delta_2$ even in the
absence of particle-hole symmetry, then the excitation energies for the $s^{+-}$ case assume the simple form 
$E_{a,\mathbf{k}}^{2}= \left(E_{a,\mathbf{k}}^{\mathrm{AFM}}\right)^{2}+\Delta^{2}$, where

\begin{equation}
E_{a,\mathbf{k}}^{\mathrm{AFM}} = \left(\frac{\xi_{1,\mathbf{k}}+\xi_{2,\mathbf{k}}}{2}\right) \pm \sqrt{M^2 + \left( \frac{\xi_{1,\mathbf{k}}-\xi_{2,\mathbf{k}}}{2} \right)^{2}} \end{equation}
are the excitation energies of the pure AFM state. 
Thus, in this special situation, one can perform two separate Bogolyubov transformations to diagonalize the full Hamiltonian. 

These considerations allow for some general conclusions of the order
parameter dependence of the free energy Eq.(\ref{F}). The quadratic
terms of $f\left(M,\Delta_{\alpha}\right)$ depend on the interaction
strengths $I$ and $\left\vert V\right\vert $. However, all other
dependencies take place only via the implicit dependence of $E_{a,\mathbf{k}}$
on the order parameters. In case of $s^{++}$ pairing, the order parameters
enter these extra dependencies through the combinations $\left(M\pm\Delta\right)^{2}$,
while for $s^{+-}$ pairing the third term in Eq.(\ref{F}) can only
depend on the combination $M^{2}+\Delta^{2}$. Thus, we find for the
free energy of the $s^{+-}$-state with particle-hole symmetry: \begin{equation}
f_{+-}=\frac{2M^{2}}{I}+\frac{2\Delta^{2}}{\left\vert V\right\vert }+2\Phi\left(M^{2}+\left\vert \Delta\right\vert ^{2}\right)\label{Free_spm}\end{equation}
On the other hand, it follows for the $s^{++}$-state: \begin{eqnarray}
f_{++}\left(\Delta,M^{2}\right) & = & \frac{2M^{2}}{I}+\frac{2\Delta^{2}}{\left\vert V\right\vert }+\Phi\left(\left(\left\vert M\right\vert +\left\vert \Delta\right\vert \right)^{2}\right)\nonumber \\
 &  & +\Phi\left(\left(\left\vert M\right\vert -\left\vert \Delta\right\vert \right)^{2}\right),\label{Free_spp}\end{eqnarray}
where \begin{equation}
\Phi\left(x\right)=-4T\rho\int_{\left\vert x\right\vert }^{\infty}dz\:\frac{z\log\left(2\cosh\left(\beta z/2\right)\right)}{\sqrt{z^{2}-x^{2}}},\end{equation}
is the same function in both cases. These facts have important implications
for the Landau expansion of the free energy that we discuss next.

In order to obtain microscopic expressions for the coefficients of
the Ginzburg-Landau theory, we expand $\delta f\left(\mathbf{M},\Delta_{\alpha}\right)=f\left(\mathbf{M},\Delta_{\alpha}\right)-f\left(0,0\right)$
with respect to $\mathbf{M}$ and $\Delta_{\alpha}$. It follows from
Eq.(\ref{F}) that

\begin{eqnarray}
\delta f & = & \frac{a_{m}}{2}\mathbf{M}^{2}+\frac{u_{m}}{4}\mathbf{M}^{4}+\sum_{\alpha,\beta}\frac{a_{s,\alpha\beta}}{2}\Delta_{\alpha}\Delta_{\beta}\nonumber \\
 &  & +\sum_{\alpha}\frac{u_{s,\alpha}}{4}\Delta_{\alpha}^{4}+\sum_{\alpha,\beta}\frac{\gamma_{\alpha\beta}}{2}\mathbf{M}^{2}\Delta_{\alpha}\Delta_{\beta}\label{free_energy_expansion}\end{eqnarray}

For the coefficients of the antiferromagnetic order parameter follows:

\begin{eqnarray}
a_{m} & = & \frac{4}{I}-2\chi_{ph}\left(\mathbf{Q}\right)\label{GL_coefficients_1}\\
u_{m} & = & \frac{1}{N}\sum_{\mathbf{k}}\frac{A_{1,\mathbf{k}}\mathrm{sech}^{2}\left(\frac{\xi_{1,\mathbf{k}}}{2T}\right)-A_{2,\mathbf{k+Q}}\mathrm{sech}^{2}\left(\frac{\xi_{2,\mathbf{k+Q}}}{2T}\right)}{T\left(\xi_{1,\mathbf{k}}-\xi_{2,\mathbf{k+Q}}\right)^{3}}\nonumber \end{eqnarray}
 with coefficients $A_{\alpha,\mathbf{k}}=-\xi_{\alpha,\mathbf{k}}+\xi_{\overline{\alpha},\mathbf{k+Q}}+2T\sinh\left(\xi_{\alpha,\mathbf{k}}/T\right)$
and the bare static particle-hole response at the antiferromagnetic
ordering vector: \begin{equation}
\chi_{ph}\left(\mathbf{Q}\right)=\frac{1}{N}\sum_{\mathbf{k}}\frac{\tanh\left(\frac{\xi_{1,\mathbf{k}}}{2T}\right)-\tanh\left(\frac{\xi_{2,\mathbf{k+Q}}}{2T}\right)}{\xi_{1,\mathbf{k}}-\xi_{2,\mathbf{k+Q}}}\end{equation}

For the coefficients of the superconducting order parameters follows
:

\begin{eqnarray}
a_{s,\alpha\beta} & = & -\frac{2}{V}\left(1-\delta_{\alpha\beta}\right)-\delta_{\alpha\beta}\chi_{pp}\left(\mathbf{0}\right)\label{GL_coefficients_2}\\
u_{s,\alpha} & = & \frac{1}{4NT}\sum_{\mathbf{k}}\frac{\mathrm{sech}^{2}\left(\frac{\xi_{\alpha,\mathbf{k}}}{2T}\right)\left(T\sinh\left(\beta\xi_{\alpha,\mathbf{k}}\right)-\xi_{\alpha,\mathbf{k}}\right)}{\xi_{\alpha,\mathbf{k}}^{3}}\nonumber \end{eqnarray}
where \begin{equation}
\chi_{pp}\left(\mathbf{0}\right)=\frac{1}{N}\sum_{\mathbf{k},\alpha}\frac{\tanh\left(\frac{\xi_{\alpha,\mathbf{k}}}{2T}\right)}{\xi_{\alpha,\mathbf{k}}}\end{equation}
is the bare static particle-particle response at external momentum
$\mathbf{q=0}$. Finally, it follows for the coefficients $\gamma_{\alpha\beta}$
that determine the coupling between both order parameters:

\begin{eqnarray}
\ \gamma_{\alpha\alpha} & = & \frac{1}{2N}\sum_{\mathbf{k,}i=1}^{3}\frac{C_{\alpha,\mathbf{k}}^{\left(i\right)}}{T\xi_{\alpha,\mathbf{k}}^{2}\left(\xi_{\alpha,\mathbf{k}}+\xi_{\bar{\alpha},\mathbf{k}}\right)\left(\xi_{\alpha,\mathbf{k}}-\xi_{\bar{\alpha},\mathbf{k}}\right)^{2}}\label{GL_coefficients_3}\\
\gamma_{\alpha\bar{\alpha}} & = & \frac{1}{N}\sum_{\mathbf{k}}\frac{\xi_{\alpha,\mathbf{k}}\tanh\left(\beta\xi_{\bar{\alpha},\mathbf{k}}/2\right)-\xi_{\bar{\alpha},\mathbf{k}}\tanh\left(\beta\xi_{\alpha,\mathbf{k}}/2\right)}{T\xi_{\alpha,\mathbf{k}}\xi_{\bar{\alpha},\mathbf{k}}\left(\xi_{\alpha,\mathbf{k}}^{2}-\xi_{\bar{\alpha},\mathbf{k}}^{2}\right)}\nonumber \end{eqnarray}
 with $C_{\alpha,\mathbf{k}}^{\left(1\right)}=2T\tanh\left(\beta\xi_{\alpha,\mathbf{k}}/2\right)\left(\xi_{\alpha,\mathbf{k}}^{2}-\xi_{\bar{\alpha},\mathbf{k}}^{2}+2\xi_{\alpha,\mathbf{k}}\xi_{\bar{\alpha},\mathbf{k}}\right)$,
$\ C_{\alpha,\mathbf{k}}^{\left(2\right)}=-\xi_{\alpha,\mathbf{k}}\left(\xi_{\alpha,\mathbf{k}}^{2}-\xi_{\bar{\alpha},\mathbf{k}}^{2}\right)\mathrm{sech}^{2}\left(\beta\xi_{\alpha,\mathbf{k}}/2\right)$
as well as $C_{\alpha,\mathbf{k}}^{\left(3\right)}=-4T\xi_{\alpha,\mathbf{k}}^{2}\tanh\left(\beta\xi_{\bar{\alpha},\mathbf{k}}/2\right)$.

These Ginzburg-Landau coefficients can also be expressed in terms
of Feynman diagrams, obtained by integrating out the fermionic degrees
of freedom of the system with competing AFM and SC. The derivation
is presented in Appendix A; in Fig. \ref{fig_feynman}, we show the
diagrammatic representation of all the quartic coefficients in terms
of the single-particle Green's functions $G_{i}\left(k\right)=\left(i\omega_{n}-\xi_{i,\mathbf{k}}\right)^{-1}$.

\begin{figure}

\begin{centering}
\includegraphics[width=1\columnwidth]{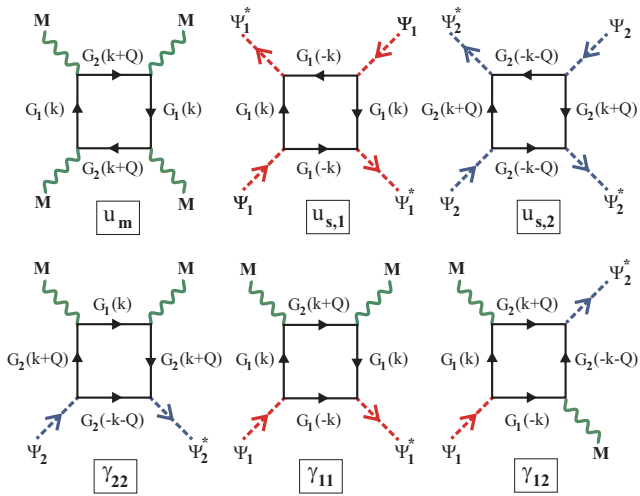} 
\par\end{centering}

\caption{Diagrammatic representation of the quartic Ginzburg-Landau coefficients
associated to the system with competing AFM and SC order parameters.
The single-particle Green's functions of the two bands are denoted
by $G_{i}\left(k\right)$. }

\label{fig_feynman} 
\end{figure}

Due to the coupling between the two bands, $\Delta_{1}$ and $\Delta_{2}$
will always appear simultaneously. As follows from the eigenvectors
of $\Lambda$ in Eq.(\ref{pairing matrix}), close to $T_{c}$, the
ratio $\Delta_{1}/\Delta_{2}=\pm\left(\rho_{2}/\rho_{1}\right)^{1/2}$
is determined by the ratio between the densities of states of the
two bands. In our case holds $\rho_{2}/\rho_{1}=\sqrt{m_{x}m_{y}}/m$.
The relative sign of $\Delta_{1}$ and $\Delta_{2}$ depends on the
sign of $V$. Thus, one can introduce the superconducting order parameter
$\Delta$ via \begin{eqnarray}
\Delta_{1} & = & \sqrt{\frac{2\rho_{2}}{\rho_{1}+\rho_{2}}}\Delta\nonumber \\
\Delta_{2} & = & \pm\sqrt{\frac{2\rho_{1}}{\rho_{1}+\rho_{2}}}\Delta\end{eqnarray}
such that $\Delta^{2}=\left(\Delta_{1}^{2}+\Delta_{2}^{2}\right)/2$.
This leads to the Landau expansion coefficients of the superconducting
order parameter of Eq.\ref{GL}\begin{eqnarray}
a_{s} & = & \frac{2a_{s,11}\rho_{2}+2a_{s,22}\rho_{1}+4\left\vert a_{s,12}\right\vert \sqrt{\rho_{1}\rho_{2}}}{\left(\rho_{1}+\rho_{2}\right)}\nonumber \\
u_{s} & = & \frac{4u_{s,1}\rho_{2}^{2}+4u_{s,2}\rho_{2}^{2}}{\left(\rho_{1}+\rho_{2}\right)^{2}}\nonumber \\
\gamma & = & \frac{2\gamma_{11}\rho_{2}+2\gamma_{22}\rho_{1}\pm4\gamma_{12}\sqrt{\rho_{1}\rho_{2}}}{\left(\rho_{1}+\rho_{2}\right)}\label{effective_GL_coeff}\end{eqnarray}

Note that the coefficient $\gamma$ of the coupling between the SC
and AFM order parameters depends on the relative phase of the two
SC order parameters. In the next section we analyze these expressions
and discuss the implications of these results for the phase diagram
of the pnictides.

Before moving on, let us briefly discuss the relationship between the system's dimensionality 
and the N\'eel transition temperature. Since our mean-field model is insensitive to fluctuations, it allows a finite $T_N$ even for two dimensions.
However, due to Hohenberg-Mermin-Wagner theorem\cite{HMW}, a non-zero $T_N$ will only be possible if the band structure has a three-dimensional dispersion, i.e. if 
the electronic interaction responsible for the AFM instability is effectively $3D$. 
In the iron arsenides, the AFM ordering involves Fe ions 
located on spatially separated layers. Assuming a weak interlayer coupling, we can introduce a phenomenological $3D$ anisotropic action for
the low-energy collective magnetic modes and obtain the N\'eel transition temperature $T_N$:

\begin{equation}
\frac{T_N - T_N^0}{T_N} \simeq \ln\left( \frac{J_z}{J} \right). \end{equation}

Here, $J$ is the effective in-plane magnetic exchange, $J_z$ is the interlayer coupling and $T_N^{0}$ is the
mean-field N\'eel transition temperature. Notice that $J$ and $J_z$ are effective parameters
of the low-energy model originated by the electronic interaction Eq. (\ref{H_AFM_full}), and are not necessarily related to localized spins. 
The logarithmic dependence of $T_{N}$ on $J_{z}/J$ shows that the overall scale of the transition temperature of an 
anisotropic magnetic material is determined by the mean-field value $T_{N}^{0}$.
This explains why, in the iron arsenides, $T_N$ has the same order of magnitude for both the 1111 and the 122 compounds, even 
though the former are much more anisotropic than the latter.

\section{Phase diagrams}

In this section, we will use the formalism developed above to explore
the possible phase diagrams of the system with competing SC and AFM
order. In particular, we will be interested in analyzing whether different
superconducting states are able to coexist with an itinerant antiferromagnetic
state or destined to phase separate from it.

\subsection{Particle-hole symmetric case}

As we stated in Section III, the band structure of the iron arsenides
can be generically described by two sets of hole and electron bands,
displaced from each other by the magnetic ordering vector $\mathbf{Q}$.
Even though the electron and hole bands are not perfectly symmetric
to each other, we can start our analysis by considering, at first,
the case of two nested bands, such that $\xi_{\mathbf{k}}\equiv\xi_{1,\mathbf{k}}=-\xi_{2,\mathbf{k+Q}}$.
Notice that, in this context, nesting does not mean that the distinct
pockets of the Fermi surface have parallel segments; instead, it implies
that they have the same shape and area, such that the non-interacting
Hamiltonian $\mathcal{H}_{0}$ has particle-hole symmetry.

The limit of perfect nesting corresponds to $\varepsilon_{0}\equiv\varepsilon_{1,0}=\varepsilon_{2,0}$,
$m_{x}=m_{y}=m$ and $\mu=0$ in Eq.(\ref{band_dispersion}). In this
case, it is straightforward to conclude that $\Delta=\left\vert \Delta_{1}\right\vert =\left\vert \Delta_{2}\right\vert $.
Moreover, using formulas (\ref{GL_coefficients_2}), it follows that
$a_{s,11}=a_{s,22}$ and $u_{s,1}=u_{s,2}$. Thus, independently of
the relative sign between the Cooper pair wave functions of the two
bands, they have the same SC Ginzburg-Landau coefficients, meaning
that the thermodynamic properties of the {}``pure'' $s^{++}$ and
$s^{+-}$ states will be the same. However, the coupling to the magnetic
degrees of freedom significantly changes this picture.

The Ginzburg-Landau expansion is formally valid around the temperature
where the AFM and SC phase lines meet, $T_{N}\simeq T_{c}$. From
the magnetic and superconducting quadratic coefficients, we see that,
for particle-hole symmetry, this condition implies $I=\left|V\right|$.
Thus, calculation of the Ginzburg-Landau coefficients using Eqs. (\ref{GL_coefficients_1}),
(\ref{GL_coefficients_2}) and (\ref{GL_coefficients_3}) yields:

\begin{eqnarray}
a & \equiv & a_{m}=a_{s}=\frac{4}{I}-\frac{2}{N}\sum_{\mathbf{k}}\frac{\tanh\left(\frac{\xi_{\mathbf{k}}}{2T}\right)}{\xi_{\mathbf{k}}}\nonumber \\
 & = & 4\left[\frac{1}{I}-\rho\:\ln\left(\frac{W}{\alpha T}\right)\right]\label{particle_hole_symm_11}\end{eqnarray}
 where $\alpha=\pi e^{-\gamma_{E}}/2$, as well as:

\begin{equation}
u\equiv u_{m}=u_{s}=\gamma_{11}=\gamma_{22}=2\gamma_{12}\label{particle_hole_symm_2}\end{equation}
 with:

\begin{eqnarray}
u & = & \frac{1}{2NT}\sum_{\mathbf{k}}\frac{\mathrm{sech}^{2}\left(\frac{\xi_{\mathbf{k}}}{2T}\right)\left(T\sinh\left(\frac{\xi_{\mathbf{k}}}{T}\right)-\xi_{\mathbf{k}}\right)}{\xi_{\mathbf{k}}^{3}}\nonumber \\
 & = & \frac{\rho}{T_{c}^{2}}\frac{7\zeta\left(3\right)}{2\pi^{2}}\label{aux_particle_hole_symm_2}\end{eqnarray}

Inserting these results into the Ginzburg-Landau expansion Eq. (\ref{free_energy_expansion})
yields\cite{Fernandes10}:\begin{eqnarray}
\delta f\left(\mathbf{M},\Delta\right) & = & \frac{a}{2}\left(\mathbf{M}^{2}+\Delta^{2}\right)+\frac{u}{4}\left(\mathbf{M}^{2}+\Delta^{2}\right)^{2}\nonumber \\
 &  & +g\:\frac{u}{2}\mathbf{M}^{2}\Delta^{2}\label{free_energy_symmetric}\end{eqnarray}
 where $g=\left(1+\cos\theta\right)$, as given by Eq. (\ref{g_parameter}),
with $\theta$ denoting the relative phase between the SC order parameters
of the electron and hole bands. Thus, for the $s^{++}$ state ($\theta=0$),
it follows that $g=2>0$, meaning that the $s^{++}$ state is deep
in the mutual exclusion regime, unable to coexist with AFM in the region of the phase diagram close to $T_N \simeq T_c$. However,
for the $s^{+-}$ state it holds that $g=0$, implying that this state
is in the borderline between the coexistence and mutual exclusion
regimes.

We can trace back to the Ginzburg-Landau coefficients in Eqs. (\ref{GL_coefficients_1}),
(\ref{GL_coefficients_2}) and (\ref{GL_coefficients_3}) the origin
for the distinct behaviors of the systems with competing itinerant
antiferromagnetism and $s^{++}$ or $s^{+-}$ superconductivity. As
we showed above, the quadratic and quartic SC and AFM coefficients
are the same in both cases. However, the net SC-AFM coupling coefficient
$\gamma_{+-}=\gamma_{11}+\gamma_{22}-2\gamma_{12}$ is reduced in
the case of an $s^{+-}$ state when compared to the case of an $s^{++}$
state, where $\gamma_{++}=\gamma_{11}+\gamma_{22}+2\gamma_{12}$.
Notice that in both situations $\gamma>0$, evidencing the competition
between the two phases.

In fact, from the diagrammatic representation of the coefficients
presented in Fig. \ref{fig_feynman}, it is clear that the only Feynman
diagram sensitive to the relative phase between the SC order parameters
of the two bands is the diagram corresponding to $\gamma_{12}$ .
It gives a contribution to the free energy of the form $M^{2}\left|\Psi_{1}\right|\left|\Psi_{2}\right|\cos\theta$, see Eq.\ref{Josephson}.
Therefore, the static long-range magnetic order plays the role of
an intrinsic Josephson coupling: it provides the momentum $\mathbf{Q}$
to the electrons of a Cooper pair in band $1$, scattering them coherently
to band $2$, where they recombine. Thus, the region of the phase
diagram where antiferromagnetism and superconductivity compete provides
an efficient tool to probe the relative phase between the Cooper pair
wave functions, an information that is usually reserved to very delicate
phase sensitive experiments\cite{Dale95,Tsuei00,Chen10}. The existence
of such a tool is even more relevant in the case of the iron arsenides,
since both $s^{++}$ and $s^{+-}$ states belong to the same irreducible
representation $A_{1g}$ of the tetragonal point group $D_{4h}$,
making interference experiments rather involved and complex\cite{Chen10,Parker09,Wu09,Linder09}.

The analysis of this limiting case with particle-hole symmetry suggests
that while the $s^{++}$ state is intrinsically unsuitable for coexistence
with the AFM phase in the iron arsenides, the $s^{+-}$ state may
or may not coexist with magnetism. In the realistic case where particle-hole
symmetry does not hold, the decision on whether the $s^{+-}$ state
is in the regime of coexistence or mutual exclusion will depend on
additional details of the band structure, as we will demonstrate in
the next subsection. This explains why some compounds present homogeneous
coexistence\cite{Ni09,Chu09,Ning09,Lester09,Laplace09,Julien09,Bernhard09,Pratt09,Christianson09},
like $\mathrm{Ba\left(Fe_{1-x}Co_{x}\right)_{2}As_{2}}$ and possibly\cite{Drew09}
$\mathrm{SmFeAs\left(O_{1-x}F_{x}\right)}$, while in others, such
as $\mathrm{LaFeAs\left(O_{1-x}F_{x}\right)}$, $\mathrm{PrFeAs\left(O_{1-x}F_{x}\right)}$,
$\mathrm{\left(Sr_{1-x}Na_{x}\right)Fe_{2}As_{2}}$, and $\mathrm{\left(Ba_{1-x}K_{x}\right)Fe_{2}As_{2}}$,
AFM and SC are mutually excluding \cite{Luetkens09,Rotundu09,Goko09,Fukazawa09,Hinkov09}.

Notice that these results do not depend on the specific functional
form of the bands dispersion relations nor on the dimensionality of
the system. They follow solely from the assumption of particle-hole
symmetry $\xi_{1,\mathbf{k}}=-\xi_{2,\mathbf{k+Q}}$. Note also that,
as we anticipated in the previous sections, the free energy (\ref{free_energy_symmetric})
for the $s^{+-}$ case is completely symmetric with respect to both
order parameters and can be characterized by the SO$(5)$ order parameter
$\overrightarrow{\mathbf{N}}=\left(\mathrm{Re}\Delta,\:\mathrm{Im}\Delta,\:\mathbf{M}\right)$.
Remarkably, a similar SO$(5)$ model has been proposed previously
for the cuprates\cite{Zhang97,Demler04}. In the context of the iron
arsenides, recent works\cite{Podolski09,Chubukov09_physC} demonstrated
that the complete, interacting particle-hole symmetric Hamiltonian
has an emergent SO$(6)$ symmetry (the other degree of freedom which
is not captured in our model is associated to an imaginary density
wave). The existence of such an emergent symmetry suggests that our
result regarding the ability of the $s^{+-}$ state to coexist with
magnetism is likely valid not only in our weak-coupling approach but
also in the strong-coupling limit.

\subsection{General case and application to $\mathrm{\mathbf{Ba\left(Fe_{1-x}Co_{x}\right)_{2}As_{2}}}$}

We now move away from the particle-hole symmetric case and consider
more specific details of the band structure dispersions of the iron
arsenides. Let us first consider small perturbations that break the
particle-hole symmetry. For instance, we first take $\varepsilon_{0}\equiv\varepsilon_{1,0}=\varepsilon_{2,0}$,
$m_{x}=m_{y}=m$, but a finite chemical potential $\mu\ll T$, i.e.
we have two detuned circular bands. An analytic expansion yields,
to leading order:

\begin{eqnarray}
g_{+-} & \approx & 0.018\left(\frac{\mu}{T}\right)^{4}\nonumber \\
g_{++} & \approx & 2+0.386\left(\frac{\mu}{T}\right)^{2}\label{expansions_detuned_bands}\end{eqnarray}

Thus, in the case of spherical detuned bands, we always find a first-order
transition between the superconducting and magnetic phases, independent
of the pairing state. This is in agreement with numerical calculations
performed by Vorontsov \emph{et al.}, which found no region of coexistence
between commensurate AFM and SC\cite{Vorontsov09}.

The second perturbation we consider is an infinitesimal ellipticity
of the electron band, such that $m_{x}=m+\delta m$ and $m_{y}=m-\delta m$,
but with $\varepsilon_{0}\equiv\varepsilon_{1,0}=\varepsilon_{2,0}$
and $\mu=0$. Such perturbation also makes $\left|\Delta_{1}\right|$
and $\left|\Delta_{2}\right|$ assume different values. In this case,
we obtain the following perturbative expansion of $g$ for the $s^{+-}$
case:

\begin{eqnarray}
g_{+-} & \approx & \left[-0.0039+0.0022\left(\frac{\varepsilon_{0}}{T}\right)^{2}\right.\nonumber \\
 &  & \left.+0.00008\left(\frac{\varepsilon_{0}}{T}\right)^{4}\right]\left(\frac{\delta m}{m}\right)^{4}\label{expansion_ellipticity}\end{eqnarray}
 while $g_{++}$ remains close to $2$. Since $\varepsilon_{0}\gg T$,
we conclude that the $s^{+-}$ state moves again to the regime of
mutual exclusion from antiferromagnetism.

\begin{figure}

\begin{centering}
\includegraphics[width=1\columnwidth]{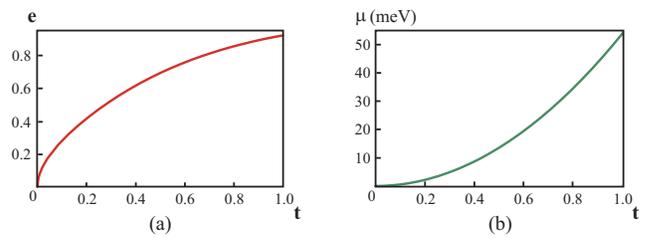} 
\par\end{centering}

\caption{(a) Ellipticity of the electron band $\mathrm{e}=\sqrt{1-\frac{m_{y}}{m_{x}}}$
and (b) chemical potential $\mu$ (in meV) as functions of $t$. The
parameter $t$ interpolates between two points of the band structure
parameters space: $t=0$ corresponds to the particle-hole symmetric
case while $t=1$ refers to the parameters that give good agreement
with experimental magnetization data\cite{Fernandes10} on $\mathrm{Ba\left(Fe_{1-x}Co_{x}\right)_{2}As_{2}}$
(see the main text for the actual values). }

\label{fig_parameter_t} 
\end{figure}

In order for the $s^{+-}$ state to be able to coexist with AFM, we
need to consider both a finite chemical potential and a finite ellipticity.
Then, depending on the particular values of the band masses, of the
energy offsets and of the chemical potential, $g_{+-}$ will be either
positive or negative, while $g_{++}$ remains positive.

To illustrate this, we perform a numerical calculation of the coefficients
$g$ through a particular path connecting two important points of
the parameters space. They are the point with particle-hole symmetry,
where $\varepsilon_{0}\equiv\varepsilon_{1,0}=\varepsilon_{2,0}$,
$m_{x}=m_{y}=m$ and $\mu=0$, and the point corresponding to the
values that consistently describe the magnetic properties of $\mathrm{Ba\left(Fe_{1-x}Co_{x}\right)_{2}As_{2}}$:
$\varepsilon_{0}=0.110$ eV, $\varepsilon_{1,0}=\varepsilon_{0}-\mu_{0}$,
$\varepsilon_{2,0}=\varepsilon_{0}+\mu_{0}$, $\mu_{0}=0.015$ eV,
$m=1.32m_{\mathrm{electron}}$, $m_{x}=2m$, $m_{y}=0.3m$ and $\mu=\mu_{c}\equiv0.039$
eV. As we showed in a previous work\cite{Fernandes10}, these parameters
give a satisfactory agreement between our model and the doping and
$T_{N}$ dependence of the experimental values of the relative zero-temperature
staggered magnetization in the absence of SC, $M\left(x,T=0\right)/M\left(x=0,T=0\right)$.
The chemical potential $\mu_{c}$ corresponds to a variation of the
electronic occupation number by $\Delta n\approx0.06$ with respect
to $\mu_{0}$. Since each added Co atom replaces one Fe atom, we associate
this increase of $\Delta n$ to the Co doping concentration $x=0.06$.

\begin{figure}

\begin{centering}
\includegraphics[width=0.8\columnwidth]{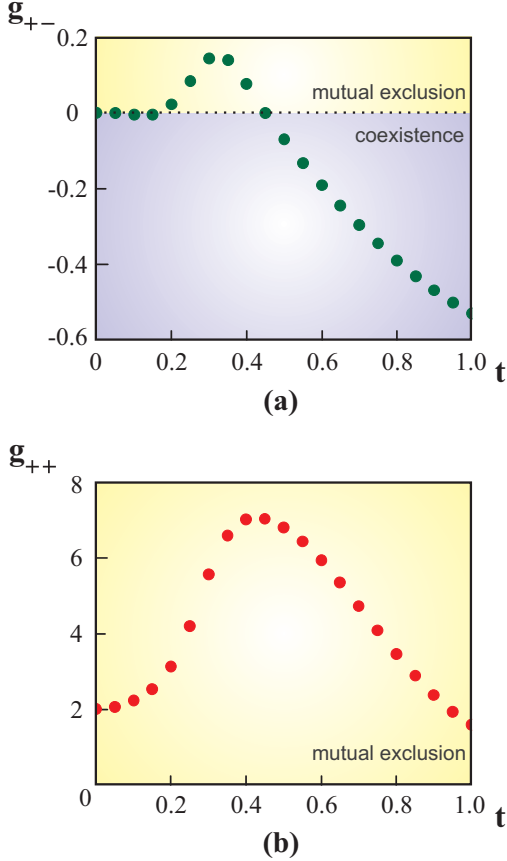} 
\par\end{centering}

\caption{Coupling coefficient $g$ for both the $s^{+-}$ state (a) and the
$s^{++}$ state (b) as function of the parameter $t$, which interpolates
between distinct band structure parameters, changing simultaneously
the electron band ellipticity and the chemical potential (see Fig.
\protect\ref{fig_parameter_t} for the definition of $t$). }

\label{fig_gpp_gpm} 
\end{figure}

In particular, the path chosen to connect these two points is parametrized
by a real number $t\in\left[0,1\right]$, such that $m_{x}=\left(1+t^{2}\right)m$,
$m_{y}=\left(1-0.7\: t\right)m$ and $\mu=\left(\mu_{c}+\mu_{0}\right)t^{2}$.
The variation of the electron band ellipticity and of the chemical
potential as function of $t$ is shown in figure \ref{fig_parameter_t},
and the corresponding values of $g_{+-}$ and $g_{++}$ are presented
in figure \ref{fig_gpp_gpm}. Clearly, when both ellipticity and $\mu$
are finite, $g_{+-}$ can be either positive or negative, but $g_{++}$
remains positive. Notice that, for the parameters corresponding to
$\mathrm{Ba\left(Fe_{1-x}Co_{x}\right)_{2}As_{2}}$ ($t=1$), the
$s^{+-}$ state coexists with magnetism, while the $s^{++}$ state
is incompatible with AFM. In particular, for $t=1$, we have:

\begin{eqnarray}
g_{+-} & \approx & -0.52\nonumber \\
g_{++} & \approx & 2.0\label{values_g_Ba122}\end{eqnarray}

\begin{figure}

\begin{centering}
\includegraphics[width=0.75\columnwidth]{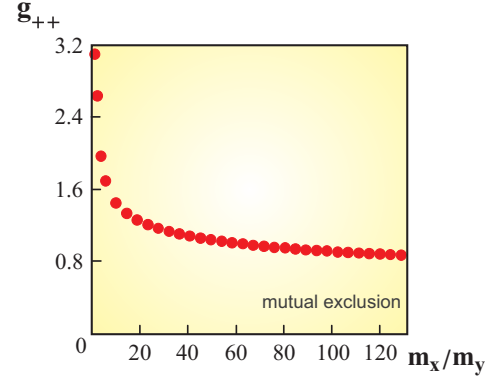} 
\par\end{centering}

\caption{Coupling coefficient $g$ for the $s^{++}$ state as function of the
electron band anisotropy $m_{x}/m_{y}$. The values of the other band
structure parameters are described in the text.}

\label{fig_gpp_mxmy} 
\end{figure}

Our analytical results and numerical calculations indicate that $g_{++}$
is generally positive, in special for the range of parameters associated
to the pnictide compounds. In order to investigate this point further,
we analyzed which parameters are able to bring $g_{++}$ to smaller
values. Particularly, we noticed that by increasing the mass anisotropy
of the electron band, while keeping the chemical potential fixed,
$g_{++}$ can be reduced. In figure \ref{fig_gpp_mxmy}, we show the
effects of an extremely large mass anisotropy on the value of $g_{++}$.
All the other band structure parameters have the values used before
for $t=1$. Clearly, even after pushing the electron band ellipticity
to unphysical limits - at least in what concerns the iron arsenides
- we still obtain that the $s^{++}$ state cannot coexist with magnetism.

These results were briefly discussed in our previous work\cite{Fernandes10}
and, in some detail, by Vavilov \emph{et al.}\cite{Vavilov09}.
Considering that $\varepsilon_{1,0},\varepsilon_{2,0}$ are the dominant
energy scales of the problem, the authors of Ref.\cite{Vavilov09}
write the band structure (\ref{band_dispersion}) in the form $\xi_{2,\mathbf{k+Q}}=-\xi_{1,\mathbf{k}}-2\delta_{\varphi}$,
where $\varphi$ is the angle along the electron pocket and $\delta_{\varphi}=\delta_{0}+\delta_{2}\cos2\varphi$,
with $\delta_{0}$ proportional to the chemical potential and band
masses difference and $\delta_{2}$ to the ellipticity. In this limit,
one can approximate $\left|\Delta_{1}\right|\approx\left|\Delta_{2}\right|$
and expand in powers of $\delta_{0}$ and $\delta_{2}$. They obtain
that $g_{+-}$ becomes negative for a significant range of
values where both $\delta_{0}$ and $\delta_{2}$ are simultaneously
finite. Furthermore, they also find that $g_{++}$ is always positive,
in complete agreement with our previous and present results.

\begin{figure}
\begin{centering}
\includegraphics[width=0.8\columnwidth]{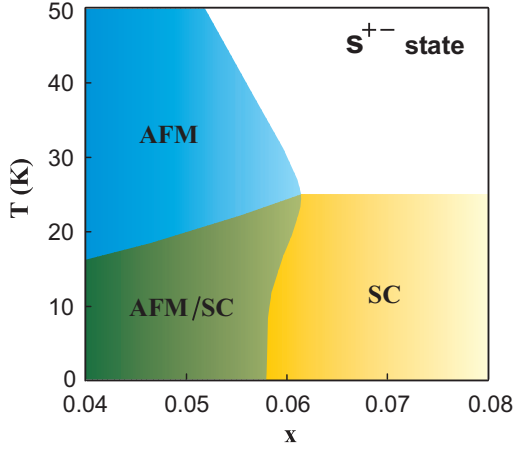} 
\par\end{centering}

\caption{Zoom of the phase diagram of $\mathrm{Ba\left(Fe_{1-x}Co_{x}\right)_{2}As_{2}}$
considering a superconducting $s^{+-}$ state (see figure \ref{fig_phase_diagrams}).
Note the reentrance of the non-magnetically ordered phase at low temperatures.
\label{fig_zoom_phase_diagram}}

\end{figure}

Although the Ginzburg-Landau expansion is extremely useful to investigate
if SC and AFM are able to coexist, it is formally not valid far from
the point where the two phase lines meet. In order to obtain a complete
$(x,T)$ phase diagram, including the back-bending of the N\'eel transition
line predicted phenomenologically in Section II, we self-consistently
solve the gap equations (\ref{gap_equations}) at a fixed occupation
number. Using the parameters discussed above for $\mathrm{Ba\left(Fe_{1-x}Co_{x}\right)_{2}As_{2}}$,
we obtain the phase diagrams presented in figure \ref{fig_phase_diagrams}.
A zoom of the phase diagram associated to the $s^{+-}$ SC state is
presented in figure \ref{fig_zoom_phase_diagram}, evidencing the
reentrance of the AFM transition line. The magnitudes of the electronic
interactions were chosen to yield\cite{Fernandes10} $T_{N}=140$
K at $x=0$ and $T_{c}=25$ K at $x=0.06$, and are given by $\left|V\right|=0.46$
eV and $I=0.95$ eV. The level of Co doping $x$ is associated to
the variation of the electronic occupation number, which depends on
the chemical potential. Specifically, we consider that each added
Co corresponds to one electron added in the system.

We emphasize that all band structure parameters were determined in
our previous work\cite{Fernandes10} by fitting the $T_{N}$ and
$x$ dependence of the experimental zero-temperature magnetization,
$M\left(x,T=0\right)/M\left(x=0,T=0\right)$, in the absence of SC.
Therefore, in the phase diagram presented in figure \ref{fig_phase_diagrams},
all the available free parameters are fixed by the shape of the transition
lines $T_{N,0}\left(x\right)$ and $T_{c,0}(x)$ of the independent,
uncoupled phases. The actual transition lines $T_{N}\left(x\right)$
and $T_{c}\left(x\right)$ of the system with coupled AFM-SC phases
are the solution of the self-consistent gap equations, with no extra
free parameters involved.

\begin{figure}
\begin{centering}
\includegraphics[width=0.8\columnwidth]{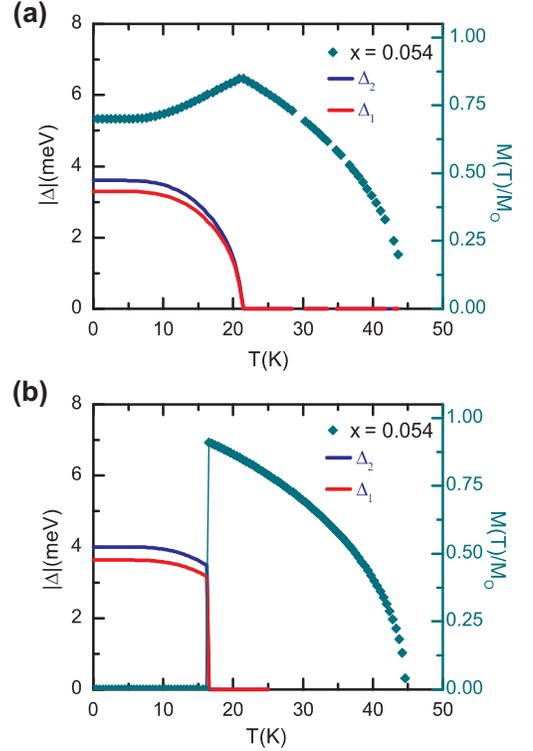} 
\par\end{centering}

\caption{Absolute values of the superconducting order parameters $\Delta_{1}$
and $\Delta_{2}$ (in meV), as well as of the magnetic order parameter
$M$ ((in units of its value $M_{0}$ at $\left(x=0,T=0\right)$))
as function of temperature $T$ (in K) for the fixed doping level
$x=0.054$ (see the $(x,T)$ phase diagrams of figure \protect\ref{fig_phase_diagrams}).
Panel (a) shows the result for an $s^{+-}$ state, whereas panel (b)
presents the result corresponding to an $s^{++}$ state. }

\label{fig_order_parameters} 
\end{figure}

\begin{figure}
\begin{centering}
\includegraphics[width=1\columnwidth]{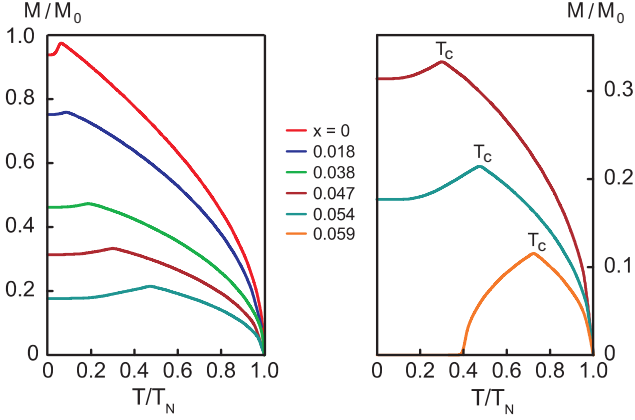} 
\par\end{centering}

\caption{Magnetic order parameter $M$ (in units of its value $M_{0}$ at $\left(x=0,T=0\right)$)
as function of temperature $T$ (in units of the AFM transition temperature
$T_{N}$) for different doping levels. Due to the competition and
coexistence with the $s^{+-}$ SC state, $M$ decreases below $T_{c}$.
Note the reentrant behavior for $x=0.059$. The right panel is a zoom
of some of the curves from the left panel. \label{fig_magnetization_curves}}

\end{figure}

Clearly, the only difference between the phase diagrams for an $s^{++}$
and an $s^{+-}$ SC state is on the coexistence / mutual exclusion
regions. In figure \ref{fig_order_parameters}, we compare the temperature
dependence of the AFM and SC gaps for a fixed doping in both cases.
For $T<T_{c}$, while in the $s^{+-}$ case the magnetization is strongly
suppressed but still survives, in the $s^{++}$ case it completely
vanishes once the SC gap opens. Note that, for the $s^{+-}$ case,
the $T=0$ values for $M$ and $\Delta_{\alpha}$ are smaller than
their values in the respective pure states. In figure \ref{fig_magnetization_curves},
we present the temperature dependence of the magnetization for several
doping values in the case of AFM competing with $s^{+-}$ SC, demonstrating
its stronger suppression as the tetracritical point is approached.

Numerical calculations of the phase diagram associated to the simplified
band structure $\xi_{2,\mathbf{k+Q}}=-\xi_{1,\mathbf{k}}-2\delta_{\varphi}$
discussed above were recently presented by Vorontsov \emph{et al.}\cite{Vorontsov10}.
Our results from figure \ref{fig_phase_diagrams} are in general agreement
with their findings. Exploring other regions of the parameters space,
they also found systems where the $s^{+-}$ coexistence region does
not persist all the way to $T=0$ as well as a small region at very
low temperatures where $s^{++}$ could in principle coexist with AFM.

A rather small region with coexistence between isotropic s-wave SC and itinerant AFM was also
found by Kato and Machida\cite{Machida88} in the context of heavy fermion compounds (see Section V for a brief discussion about these materials).
Considering a single band without particle-hole symmetry, they performed numerical calculations to determine the phase diagram for different pairing states.
In particular, coexistence between isotropic s-wave SC and AFM was only found far from the multicritical point $T_N \simeq T_c$ and in a very narrow regime,
analogous to what was reported by Vorontsov \emph{et al.}\cite{Vorontsov10} in the context of the iron arsenides. Note that these results are not
in contradiction to our conclusions, since our Ginzburg-Landau expansion - and, consequently, the definition of the coupling parameter $g_{++}$ - is only valid for $T_N \simeq T_c$. 
Far from the multicritical point and from particle-hole symmetry, the details of the bands dispersions are very important
and it is in principle possible to find coexistence even if $g_{++}>0$ at $T_N \simeq T_c$.

So far, we have only compared the $s^{+-}$ and $s^{++}$ SC states in
our calculations. However, electronic theories for the superconductivity
in the iron arsenides have also proposed other symmetries for the
Cooper pair wave function where nodes are present \cite{Graser09,Chubukov09_nodes,Thomale09}.
One example is the $d$-wave state, where $\Delta_{1}$ and/or $\Delta_{2}$
have nodes along their respective Fermi pockets. The generalization
of our formalism to these other symmetry states is straightforward.
One has only to introduce the corresponding angular factors $\eta\left(\varphi\right)$
for the gaps and for the pairing interaction $V$, and then average
over the Fermi pockets.

In the case of particle-hole symmetry, where the free energy is given
by Eq. (\ref{free_energy_symmetric}), we obtain, for a $d$-wave
state, $g=\left(\sqrt{\frac{8}{3}}-1\right)u\approx0.6u$. Unlike
the $s^{+-}$ state, the $d$-wave state is not on the borderline
between coexistence and mutual exclusion from AFM. However, it is
neither deep in the mutual exclusion regime, as the $s^{++}$ state
is. Even though the $s^{+-}$ state is the most compatible with itinerant
magnetism, we cannot exclude that the $d$-wave state is also able
to coexist with AFM for certain parameters.

\subsection{Incommensurability and the sign of the coupling coefficient}

Our model assumes that the magnetism is commensurate. Experimentally,
this is still an unsettled issue for $\mathrm{Ba\left(Fe_{1-x}Co_{x}\right)_{2}As_{2}}$:
while neutron diffraction measurements did not detect any incommensurability
inside their resolution window \cite{Pratt09,Lester09,Fernandes10},
some works employing NMR\cite{Laplace09} and M\"ossbauer spectroscopy\cite{Bonville10}
suggest that the magnetism could be weakly incommensurate in these
systems.

Theoretically, the weak-coupling model for the excitonic itinerant
magnetism naturally predicts the onset of an incommensurate AFM state
for small enough temperatures\cite{Rice70}, as recently pointed
out by Vorontsov \emph{et al}\cite{Vorontsov09}. To see how this
comes out from the model we used here, consider the specific case
of detuned bands having the same shape, $\xi_{2,\mathbf{k+Q}}=-\xi_{1,\mathbf{k}}-2\mu$.
Instead of expressing the Ginzburg-Landau coefficients as momentum
sums, Eqs. (\ref{GL_coefficients_1}), (\ref{GL_coefficients_2})
and (\ref{GL_coefficients_3}), we can equivalently express them as
Matsubara sums. Using the diagrammatic form of the coefficients (see
figure \ref{fig_feynman} and Appendix A), this is a straightforward
calculation and yields:

\begin{eqnarray}
u_{m} & = & 4\pi\rho T\sum_{\omega_{n}>0}\frac{\omega_{n}\left(\omega_{n}^{2}-3\mu^{2}\right)}{\left(\omega_{n}^{2}+\mu^{2}\right)^{3}}\nonumber \\
u_{s,\alpha} & = & 2\pi\rho T\sum_{\omega_{n}>0}\frac{1}{\omega_{n}^{3}}\nonumber \\
\gamma_{\alpha\alpha} & = & 4\pi\rho T\sum_{\omega_{n}>0}\frac{\omega_{n}}{\left(\omega_{n}^{2}+\mu^{2}\right)^{2}}\nonumber \\
\gamma_{\alpha\bar{\alpha}} & = & 2\pi\rho T\sum_{\omega_{n}>0}\frac{1}{\omega_{n}\left(\omega_{n}^{2}+\mu^{2}\right)}\label{GL_coefficients_Matsubara}\end{eqnarray}

Similar expressions were obtained in Ref.\cite{Vavilov09}. Using
Eq. (\ref{GL_coefficients_Matsubara}), it becomes clear now that
$u_{m}<0$ for $T_{m}^{\ast}\lesssim0.5\mu$, indicating that the
transition from the paramagnetic to the commensurate AFM phase is
first order. However, as discussed elsewhere\cite{Vorontsov09},
when this condition is met an incommensurate AFM phase has a lower
energy than the commensurate state. For the parameters we used to
obtain the phase diagrams of $\mathrm{Ba\left(Fe_{1-x}Co_{x}\right)_{2}As_{2}}$
(Fig. \ref{fig_phase_diagrams}), where the Fermi pockets have actually
different shapes, the AFM phase line meets the SC phase line before
this incommensurate instability takes place. Even if they met after
this instability point, it was shown by Vorontsov \emph{et al.} that
the $s^{+-}$ state remains able to coexist with an incommensurate
antiferromagnetic state\cite{Vorontsov09,Vorontsov10}.

Not only does $u_{m}$ become negative for low temperatures, but also
the net coupling coefficient $\gamma\equiv\gamma_{11}+\gamma_{22}\pm2\gamma_{12}$.
Using Eqs. (\ref{GL_coefficients_Matsubara}) for detuned circular
bands, we obtain for the $s^{+-}$ case:

\begin{equation}
\gamma_{+-}=4\pi\rho\left(\varepsilon_{F}\right)T\sum_{\omega_{n}>0}\frac{\omega_{n}^{2}-\mu^{2}}{\omega_{n}\left(\omega_{n}^{2}+\mu^{2}\right)^{2}}\label{aux_GL_coefficients_Matsubara}\end{equation}

Thus, $\gamma_{+-}<0$ for $T^{\ast}\lesssim0.3\mu$. Although the
sign of $\gamma_{+-}$ does not affect the criterion for phase coexistence,
$\gamma^{2}<u_{m}u_{s}$, it significantly changes the forms of the
AFM and SC transition lines inside the coexistence region. In particular,
a negative $\gamma$ implies that neither $T_{c}$ nor $M$ are suppressed
in the AFM-SC coexistence regime - see, for instance, Eq. (\ref{condition_suppression}).

A similar result for the AFM-SC coupling coefficient $\gamma$ was
obtained by Zhang \emph{et al.} in the context of the cuprates\cite{Sachdev02}.
In a weak-coupling calculation at $T=0$ but finite disorder (otherwise
the Matsubara sums would diverge), they obtain a negative coupling
coefficient between a single-band $d$-wave SC order parameter and
an itinerant AFM order parameter. Technically, the problem of the
competition between AFM and a single-band $d$-wave SC is equivalent
to our two-band problem with the $s^{+-}$
SC state. Notice, however, that the coefficient $\gamma_{+-}$ only
becomes negative at $T^{\ast}<T_{m}^{\ast}$, i.e. the incommensurate
AFM transition would happen before the coupling coefficient changes
sign. Thus, $\gamma_{+-}$ has no meaning in this regime and one would
have to go back and calculate the coupling coefficient between an
incommensurate AFM order parameter and the SC order parameter. However,
the numerical calculations performed by Vorontsov \emph{et al.} indicate that 
this coupling coefficient must be positive\cite{Vorontsov09,Vorontsov10}.
Thus, in our approach, the most stable AFM state and superconductivity are always competing.

Notice that this theoretical discussion about the incommensurability
of the AFM state does not take into account the coupling to the lattice
degrees of freedom. As argued by many authors, an emergent nematic
degree of freedom is present in the iron arsenides due to its frustrated
magnetic structure\cite{Fang08,Xu08,Fernandes09}. The energy of
the system is minimized by the onset of a nematic transition at $T_{\mathrm{nem}}\geq T_{N}$.
Due to the bilinear coupling between the nematic order parameter and
the shear distortion, the nematic transition is simultaneous to a
structural transition from the tetragonal to the orthorhombic phase.
Key to this process is the commensurability of the magnetic fluctuations
that give rise to this emergent nematic degree of freedom\cite{Fernandes09}.
Thus, the inclusion of this extra degree of freedom could change the
outcome of an incommensurate AFM state at low temperatures.

\subsection{Intraband pairing and Coulomb interaction}

In writing our weak-coupling expression for the SC interaction term,
Eq. (\ref{H_SC}), we considered only an interband pairing interaction
$V\equiv V_{12}=V_{21}$. If one includes additional intra-band pairing
interactions $V_{11}$ and $V_{22}$, the only change in the free
energy density, Eq.(\ref{F}), is that the quadratic term $\ -\frac{1}{V}\left(\Delta_{1}^{\ast}\Delta_{2}+\Delta_{2}^{\ast}\Delta_{1}\right)$
is replaced by $-\sum_{\alpha\beta}\left(V^{-1}\right)_{\alpha\beta}\Delta_{\alpha}^{\ast}\Delta_{\beta}.$
This will of course change the gap equations, specially the value
of $T_{c}$, and may also affect the ratio $\Delta_{1}/\Delta_{2}$.
Yet, the inclusion of an intraband pairing interaction will only change
the quadratic Ginzburg-Landau coefficients $a_{s,\alpha}$, leaving
the values of the quartic coefficients $u_{s}$, $u_{m}$ and $\gamma$
unchanged. Since our results regarding the coexistence or mutual exclusion
between the SC and AFM states rely solely on the quartic coefficients,
they will remain unchanged.

Here, we assume that the interband pairing interaction $V\equiv V_{12}$
is originated from the coupling between electrons and collective modes
of the system, such as phonons ($V<0$) or paramagnons ($V>0$), for
example. With this in mind, we can investigate the effects of the
electronic Coulomb repulsion by adding a renormalized Coulomb interaction
$U>0$. First, consider the case of an uniform
Coulomb repulsion, with equal intraband and interband terms $U$.
Formally, there is now a single interband interaction given by $V+U$,
which is enhanced (reduced) in the case of $s^{+-}$ ($s^{++}$) pairing.
Yet, due to the different origins of $V$ and $U$, we here opt to
write the total interband interaction in the form $V+U$.

For the pure $s^{+-}$ state, it was previously shown that an uniform
renormalized Coulomb interaction $U$ is unable to completely destroy
the SC state, i.e. $T_{c}\left(U\right)$ never goes to zero, no matter
the magnitude of the Coulomb interaction\cite{Mazin09}. In order
to demonstrate this, one writes the linearized gap equations in matrix
form as $\Delta_{\alpha}=\Lambda_{\alpha\beta}\Delta_{\beta}$ and
analyzes the eigenvalues of:\begin{equation}
\Lambda=\left(\begin{array}{cc}
-U\rho_{1} & -\left(V+U\right)\rho_{2}\\
-\left(V+U\right)\rho_{1} & -U\rho_{2}\end{array}\right)\label{linearized_gap_coulomb}\end{equation}

\begin{figure}

\begin{centering}
\includegraphics[width=0.8\columnwidth]{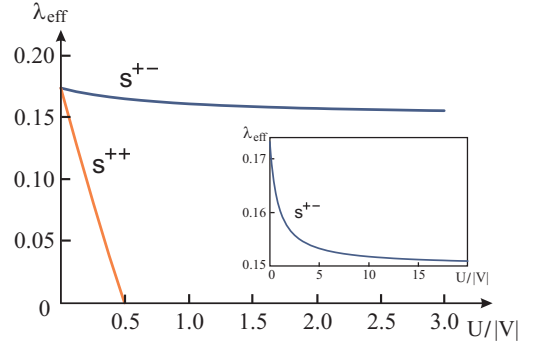} 
\par\end{centering}

\caption{Effective SC coupling constant $\lambda_{\mathrm{eff}}=1/\ln\left(W/\alpha T_{c}\right)$
as function of the ratio between the Coulomb repulsion and the pairing
interaction $U/\left|V\right|$ for both a pure $s^{++}$ state and
a pure $s^{+-}$ state. Here, we considered the values $\left\vert V\right\vert \rho_{1}=0.1$
and $\left\vert V\right\vert \rho_{2}=0.3$, but the conclusions are
similar for arbitrary parameters. The inset is a zoom of the curve associated to the $s^{+-}$ state.}

\label{fig_coulomb_repulsion} 
\end{figure}

The largest eigenvalue $\lambda$ determines the transition temperature
through $\lambda^{-1}=\ln\left(W/\alpha T_{c}\right)$. For small
$U$, it follows that $\lambda=\left\vert V\right\vert \sqrt{\rho_{1}\rho_{2}}-\frac{1}{2}\left(\sqrt{\rho_{1}}\pm\sqrt{\rho_{2}}\right)^{2}U$,
where $+1$ refers to $s^{++}$ pairing and $-1$ to $s^{+-}$ pairing,
respectively. The suppression of the pairing interaction is significantly
weaker for the $s^{+-}$-state, in particular for similar densities
of states $\rho_{1}$ and $\rho_{2}$. For $s^{++}$-pairing $\lambda$,
and thus $T_{c}$, vanishes as $U\rightarrow\left\vert V\right\vert /2$,
while in case of $s^{+-}$-pairing the net pairing interaction stays
finite even for infinite $U$, where it holds $\lambda\left(U\rightarrow\infty\right)=2\left|V\right|\rho_{1}\rho_{2}/\left(\rho_{1}+\rho_{2}\right)$.
These results are summarized in figure \ref{fig_coulomb_repulsion}.

We next investigate the effects of an uniform Coulomb repulsion in
the case where $s^{+-}$-SC coexists with AFM. We now have:

\begin{equation}
\Lambda=\left(\begin{array}{cc}
-\left(V+U\right)r_{0}-Ur_{2} & -\left(V+U\right)r_{1}-Ur_{0}\\
-\left(V+U\right)r_{2}-Ur_{0} & -\left(V+U\right)r_{0}-Ur_{1}\end{array}\right)\label{linearized_gap_eq}\end{equation}
 where

\begin{eqnarray}
r_{\alpha} & = & \frac{1}{N}\sum_{\mathbf{k},a}\frac{\left(E_{a,\mathbf{k}}^{2}-\xi_{\alpha,\mathbf{k}}^{2}\right)\tanh\left(\beta E_{a,\mathbf{k}}/2\right)}{2E_{a,\mathbf{k}}\left(E_{a,\mathbf{k}}^{2}-E_{\bar{a},\mathbf{k}}^{2}\right)}\nonumber \\
r_{0} & = & \frac{1}{N}\sum_{\mathbf{k},a}\frac{M^{2}\tanh\left(\beta E_{a,\mathbf{k}}/2\right)}{2E_{a,\mathbf{k}}\left(E_{a,\mathbf{k}}^{2}-E_{\bar{a},\mathbf{k}}^{2}\right)}\label{aux_linearized_gap_eq}\end{eqnarray}
 and the excitation energies are given by Eq. (\ref{excitation_energies})
with $\Delta_{1}=\Delta_{2}=0$. The superconducting transition temperature
is again determined by the largest eigenvalue of $\Lambda$, which
is given by:

\begin{eqnarray}
\lambda & = & -\left(r_{1}+r_{2}+2r_{0}\right)U-2r_{0}V+\left[\left(r_{1}+r_{2}+2r_{0}\right)^{2}U^{2}\right.\nonumber \\
 &  & \left.+4r_{1}r_{2}V^{2}+4UV\left(2r_{1}r_{2}+r_{0}r_{1}+r_{0}r_{2}\right)\right]^{1/2}\label{largest_eigenvalue}\end{eqnarray}

For $U=0$, we find $\lambda=2\left(\sqrt{r_{1}r_{2}}-r_{0}\right)V$.
Let us assume that, in the absence of the Coulomb interaction, the
system undergoes a SC transition at $T_{c}$. Imposing the vanishing
of the largest eigenvalue, we obtain that $\lambda\left(U_{c}\right)=0$
for $U_{c}=-V/2$, independent of the value of the magnetic order
parameter $M$ or of band structure details. However, for the $s^{+-}$
case, $V>0$, implying $U_{c}<0$. Therefore, the $s^{+-}$ SC state
inside the antiferromagnetic phase is robust against an uniform Coulomb
interaction, similarly to what happens for the nonmagnetic $s^{+-}$
state.

We can also consider the case where the Coulomb interaction is not
uniform, such that its value for intraband repulsion $U$ is greater
than its value for interband repulsion $U^{\prime}<U$. Now we need
to determine the largest eigenvalue $\lambda$ of

\begin{equation}
\Lambda=\left(\begin{array}{cc}
-\left(V+U^{\prime}\right)r_{0}-Ur_{2} & -\left(V+U^{\prime}\right)r_{1}-Ur_{0}\\
-\left(V+U^{\prime}\right)r_{2}-Ur_{0} & -\left(V+U^{\prime}\right)r_{0}-Ur_{1}\end{array}\right)\label{linearized_gap_eq_prime}\end{equation}

$\lambda$ now vanishes for $U=\left\vert V\right\vert \pm U^{\prime}$,
where the plus (minus) sign is to be used for the case of an $s^{+-}$
($s^{++}$) state. In this case, both states can be destroyed by a
sufficiently large repulsion, yet the $s^{++}$ state is destroyed
easier than the $s^{+-}$ state. Note also that the condition $U=\left\vert V\right\vert \pm U^{\prime}$
is the same for both situations of a pure SC state and a coexistent
SC-AFM state. Thus, in general, magnetism does not seem to significantly
influence the ability of the Coulomb repulsion to destroy the SC order.
The renormalization of the Coulomb interaction was also investigated
using a renormalization group approach in Ref.\cite{Chubukov08}.
Even though the underlying reasoning is somewhat different from our
analysis, the conclusions of Ref.\cite{Chubukov08} are consistent
with our result. The pair-breaking contribution of the Coulomb interaction
is less efficient for the $s^{+-}$-state, if compared to $s^{++}$-superconductivity.

\subsection{Reentrant N\'eel transition line and quantum fluctuations}

According to our phenomenological discussion in Section II, the strong
suppression of the AFM order parameter in the SC phase is also reflected
in the reentrance of the AFM transition line, as shown by the calculated
phase diagrams of figure \ref{fig_phase_diagrams} and confirmed by
neutron diffraction measurements\cite{Fernandes10}. The same reentrant
behavior is observed in some heavy fermions\cite{Knebel09,Isaacs95},
where AFM-SC coexistence takes place as well. In the cuprates, theoretical
models also proposed that a similar reentrance is present in the phase
diagram\cite{Sachdev10_arxiv}.

Let us investigate in more detail the form of the reentrance line
in the iron arsenides. For simplicity, we follow Ref.\cite{Vavilov09}\emph{\ }and
consider a small perturbation of the particle-hole symmetric band
structure, $\xi_{2,\mathbf{k+Q}}=-\xi_{1,\mathbf{k}}-2\delta_{\varphi}$,
with $\delta_{\varphi}=\delta_{0}+\delta_{2}\cos2\varphi$, as we
explained in Section IV-B. Assuming that $\delta_{0}$ and $\delta_{2}$
satisfy the conditions for coexistence between AFM and $s^{+-}$-SC,
we expand the free energy (\ref{F}) only in powers of $M$, keeping
the SC gap $\Delta=\Delta_{1}=-\Delta_{2}$ fixed. This last assumption
is justified at low temperatures, where the SC order parameter saturates
(see Fig. \ref{fig_order_parameters}). For $T\ll\delta\ll\Delta$,
we obtain\ for the quadratic magnetic coefficient:

\begin{equation}
a_{m}\left(\Delta,T\right)\approx4\left(\frac{\Delta-\Delta_{c}}{\Delta_{c}}\right)-\left(4\sqrt{2\pi}\frac{\Delta^{2}\sqrt{T}}{\delta^{2}\sqrt{\Delta}}\right)\mathrm{e}^{-\frac{\Delta}{T}}\label{am_Delta_finite}\end{equation}
with $\delta^{2}\equiv\left\langle \delta_{\varphi}^{2}\right\rangle =\delta_{0}^{2}+\delta_{2}^{2}/2$.
Here, $\Delta_{c}$ corresponds to the $T=0$ value of the SC gap
where the quantum phase transition from the superconducting normal
state to the superconducting state with antiferromagnetic long range
order takes place. Eq. (\ref{am_Delta_finite}) implies that, at $T=0$,
there is AFM order for $\Delta<\Delta_{c}$. The negative sign in
front of the temperature dependent term also implies that $T_{N}$
is finite for $\Delta>\Delta_{c}$. Therefore, it correctly captures
the reentrance of the AFM line.

The presence of the exponential term $\mathrm{e}^{-\frac{\Delta}{T}}$
is a consequence of the fact that, inside the SC state, quasiparticle
excitations are fully gapped. Due to this exponential dependency,
the reentrant $T_{N}$ line approaches the quantum critical point
with an exponentially steep slope, i.e. as an almost vertical line.
This is in agreement with our calculated phase diagrams from figure
\ref{fig_phase_diagrams}, as well as with the phase diagrams obtained
by Vorontsov \emph{et al}\cite{Vorontsov10}.

These results were derived using a mean-field approach. Close to $T=0$,
the presence of quantum fluctuations change this scenario. To illustrate
their effects, we consider collective magnetic fluctuations in the
vicinity of this quantum critical point. In order to properly describe
long range mangnetic order we have to include interlayer coupling
and consider an effective three dimensional quantum rotor model. Due
to the fact that quasiparticle excitations are gapped, we consider
a rotor model with dynamic critical exponent\cite{Sachdev_book}
$z=1$. Within a self consistent large-$N$ theory, where $N$ refers
to the number of components of the rotor, we obtain a renormalization
of the coefficient in $a_{m}\rightarrow\tilde{a}_{m}$ due to critical
fluctiations:

\begin{equation}
\tilde{a}_{m}=a_{m}+u_{m}T\sum_{\omega_{n}}\int\frac{d^{3}q}{\left(2\pi\right)^{3}}\frac{1}{\tilde{a}_{m}+q^{2}+\omega_{n}^{2}}\label{quantum_corrections}\end{equation}

An expansion at low temperatures yields $\tilde{a}_{m}-a_{m}=CT^{2}$,
with $C>0$, which dominates over the exponential term $\mathrm{e}^{-\frac{\Delta}{T}}$
that follows from the mean field theory. The presence of power-law
corrections in $a_{m}$ is more general than our self-consistent large-$N$
theory, and is expected to occur due to the presence of massless critical
fluctuations. Due to the fact that the coefficient $C$ is positive,
such quantum fluctuations suppress the magnetic reentrant behavior
at very low temperatures. However, as usual in systems in the weak
coupling regime, the critical region where quantum fluctuations are
relevant is expected to be very small, and probably hard to be detected
experimentally.

\section{Localized \emph{versus} itinerant magnetism}

A key conclusion of our calculation is that homogeneous coexistence
of superconductivity and magnetism is only allowed in case of unconventional
$s^{+-}$-pairing state, while both ordered states exclude each other
in case of conventional $s^{++}$-pairing. This conclusion seems to
be at odds with the well known fact that antiferromagnetism and superconductivity
do coexist homogeneously in a number of materials where the evidence
for conventional electron-phonon pairing is very strong, such as the
borocarbides\cite{Canfield98} $\mathrm{RNi_{2}B_{2}C}$ and the
ternary superconductors\cite{Maple82} $\mathrm{RMo_{6}S_{8}}$
and $\mathrm{RRh_{4}B_{4}}$, with $\mathrm{R}$ denoting a rare
earth. The crucial difference between these rare earth based systems
and the iron pnictide superconductors is that the magnetism in the
former is due to localized rare earth spins while in the latter the
same electrons that superconduct are responsible for the entire ordered
moment. Thus, for our argumentation in the pnictides to hold, it is
essential that the same electrons that form the Cooper pair condensate
are responsible for the ordered moment. This is evident from our Hamiltonian,
Eq.(\ref{full_Hamiltonian}), where the order parameters $\Delta$
and $M$ are expectation values of electronic states of the same bands.
This is the reason for the highly symmetric form of the free energy
of Eqs.(\ref{Free_spm},\ref{Free_spp}) and why the Ginzburg-Landau
coefficients for the quartic magnetic, superconducting and coupling
terms are closely related to each other.

In order to demonstrate explicitly that the phase diagram of competing
magnetism and superconductivity is very different in case of localized
spins, here we analyze this problem in some detail. We recall that
the total Hamiltonian is given by $\mathcal{H}_{0}+\mathcal{H}_{\mathrm{AFM}}+\mathcal{H}_{\mathrm{SC}}$.
We keep the same terms for the kinetic and superconducting parts,
given by Eqs. (\ref{H0}) and (\ref{H_SC}), respectively. The pairing
interaction $V$ in $\mathcal{H}_{\mathrm{SC}}$ might, for example,
be due to the electron-phonon interactions. At this point we are not
concerned whether systems like the $\mathrm{RNi_{2}B_{2}C}$ are
indeed characterized by a corresponding two band model. Instead, we
are primarily interested in comparing localized and itinerant magnetism
for a system with otherwise unchanged electronic structure. It will
become evident below that our analysis is in fact more general. The
crucial new term in the Hamiltonian is $\mathcal{H}_{\mathrm{AFM}}$
which is replaced by:

\begin{equation}
\mathcal{H}_{\mathrm{AFM}}=\frac{J_{K}}{4}\sum_{i}\mathbf{S}_{i}\cdot\left(c_{is}^{\dagger}\boldsymbol{\sigma}_{ss^{\prime}}d_{is^{\prime}}+h.c.\right).\label{H_AFM_local}\end{equation}

Here $\mathbf{S}_{i}$ refers to a localized spin-$S$ operator and
$J_{K}$ is the exchange coupling between localized spins and conduction
electrons. We are interested in the regime where $J_{K}$ leads to
magnetic long range order via the RKKY mechanism with $J_{\mathrm{RKKY}}\left(\mathbf{r}\right)\simeq J_{K}^{2}\chi_{s}\left(\mathbf{r}\right)$,
where $\chi_{s}\left(\mathbf{r}\right)$ denotes the electronic spin
susceptibility. In the regime of antiferromagnetism with large ordered
local moments, it is possible to neglect the Kondo effect as $J_{\mathrm{RKKY}}$
is larger than the corresponding Kondo temperature. To proceed, we
perform a mean field analysis of this model. We introduce the expectation
values:

\begin{eqnarray}
\left\langle S_{i}^{z}\right\rangle  & = & m_{\mathrm{loc}}\mathrm{e}^{i\mathbf{Q}\cdot\mathbf{R}_{i}}\nonumber \\
\left\langle s_{i}^{z}\right\rangle  & \equiv & \frac{1}{2N}\sum_{\mathbf{p}\sigma}\sigma\left\langle c_{\mathbf{p}\sigma}^{\dagger}d_{\mathbf{p}+\mathbf{Q}\sigma}\right\rangle =-m_{\mathrm{el}}\mathrm{e}^{i\mathbf{Q}\cdot\mathbf{R}_{i}}\label{order_parameters_local}\end{eqnarray}
 with magnetic ordering vector $\mathbf{Q}$. For definiteness, we
consider $J_{K}>0$, implying that $\left\langle S_{i}^{z}\right\rangle $
and $\left\langle s_{i}^{z}\right\rangle $ have opposite sign. Since
we ignore the Kondo effect, our final results are independent on the
sign of $J_{K}$.

In analogy to the theory of itinerant magnetism we perform a mean
field calculation, giving rise to the total free energy density $F=F_{\mathrm{s}}+F_{0,\mathrm{el}}+F_{\mathrm{sc}}$
with contributions from localized spins, $F_{\mathrm{s}}$, from the
SC condensate $F_{\mathrm{sc}}$, and from the electronic part, $F_{0,\mathrm{el}}$,
which also includes the order parameters coupling. The last two terms
are completely analogous to the case of an itinerant AFM state competing
with SC, Eq. (\ref{F}), if we identify the magnetic order parameter
as:

\begin{equation}
M=\frac{J_{K}m_{\mathrm{loc}}}{4}.\label{SDW_gap_localized}\end{equation}

Recall that $M$ in our notation is the antiferromagnetic potential
that causes a gap for Bragg reflected points of the Fermi surface.
In case of itinerant magnetism this gap is due to the electron-electron
interaction $I$ and the moment of the itinerant electrons. Now the
microscopic origin of $M$ is very different. Yet, the expression
for the energy of the conduction electrons is still given by:

\begin{equation}
F_{0,\mathrm{el}}=-2T\sum\limits _{\mathbf{k},a}\ln\left[2\cosh\left(\frac{E_{a,\mathbf{k}}}{2T}\right)\right],\label{F_el_local}\end{equation}
with the same excitation energies $E_{a,\mathbf{k}}=E_{a,\mathbf{k}}\left(\Delta_{1},\Delta_{2},M\right)$
from Eq. (\ref{excitation_energies}). The contribution to the energy
due to the pairing interaction is unchanged as well and given by

\begin{equation}
F_{\mathrm{sc}}=-\sum_{\alpha\beta}V_{\alpha\beta}^{-1}\Delta_{\alpha}^{\ast}\Delta_{\beta}.\label{F_SC_local}\end{equation}

Finally, the free energy density due to the localized spins is:

\begin{equation}
F_{\mathrm{s}}=-T\:\ln\sum_{m=-S}^{S}e^{m\beta h}+J_{K}m_{\mathrm{loc}}m_{\mathrm{el}},\label{F_S_local}\end{equation}
where $h=J_{K}m_{\mathrm{el}}$ is the Weiss field of a single spin-$S$.
Since both magnetizations $m_{\mathrm{loc}}$ and $m_{\mathrm{el}}$
order simultaneously, we can eliminate $m_{\mathrm{el}}$ and express
the Landau expansion in terms of $M$. To this end we use $m_{\mathrm{loc}}=-\left.\partial F_{\mathrm{s}}/\partial h\right\vert _{h=J_{K}m_{\mathrm{el}}}$
and solve for $m_{\mathrm{el}}=m_{\mathrm{el}}\left(m_{\mathrm{loc}}\right)$.
Using $M$ of Eq.\ref{SDW_gap_localized} instead of $m_{\mathrm{loc}}$,
we find to leading order \begin{equation}
m_{\mathrm{el}}=\frac{T}{\alpha_{S}J_{K}^{2}}\left(M+\frac{\beta_{S}}{4\alpha_{S}^{3}J_{K}^{2}}M^{3}\right),\end{equation}
with \begin{eqnarray}
\alpha_{S} & = & \frac{S\left(S+1\right)}{12},\nonumber \\
\beta_{S} & = & \frac{S\left(S+1\right)\left[1+2S\left(S+1\right)\right]}{90}.\end{eqnarray}

After inserting this result for $m_{\mathrm{el}}$ into the free energy,
Eq.\ref{F_S_local}, we can expand it and thus determine the Ginzburg-Landau
expansion simultaneously for the SC and AFM order parameters, $\Delta_{1}$,
$\Delta_{2}$ and $M$. We obtain the exact same expressions for the
coefficients related to the superconducting order parameter, $a_{s,\alpha\beta}$
and $u_{s,\alpha}$, as well as for the coupling $\gamma_{\alpha\beta}$
between the AFM and SC order parameters. Despite the same formal expression,
the physical interpretation of the order parameter coupling terms
$\gamma_{\alpha\beta}$ is somewhat different now. It reflects changes
in the conduction-electron mediated RKKY interaction due to the onset
of superconductivity.

The only difference in the values of the Ginzburg-Landau parameters
due to the presence of localized spins is for the coefficients of
the magnetic order parameter. We find: \begin{equation}
a_{m}=\frac{4T}{\alpha_{S}J_{K}^{2}}-2\chi_{ph}\left(\mathbf{Q}\right)\end{equation}
 and \begin{equation}
u_{m}=u_{m}^{0}+\frac{\beta_{S}T_{N}}{\left(\alpha_{S}J_{K}\right)^{4}}.\label{um loc}\end{equation}
 where $u_{m}^{0}$ is the quartic coupling of itinerant spins given
by Eq(\ref{GL_coefficients_1}). The additional term in Eq. (\ref{um loc})
is solely determined by the N\'eel temperature, $T_{N}$, the size of
the spin, $S$, and the coupling $J_{K}$. $T_{N}$ is determined
via $a_{m}\left(T_{N}\right)=0$ and given, as expected, by the RKKY
coupling\begin{equation}
T_{N}=\frac{1}{2}\:\alpha_{S}J_{K}^{2}\chi_{ph}\left(\mathbf{Q}\right)\ \end{equation}
 with bare spin susceptibility of the conduction electrons at the
ordering vector $\mathbf{Q}$. Since Eq. (\ref{aux_particle_hole_symm_2})
gives $u_{m}^{0}\approx\rho/\left(2T_{N}^{2}\right)$ , it follows
that the relative change in the magnetic Ginzburg-Landau coefficient
is: \begin{equation}
\frac{u_{m}-u_{m}^{0}}{u_{m}^{0}}\approx\frac{2\beta_{S}}{\alpha_{S}}\left(J_{K}\rho\right)^{2}\ln^{3}\left(\alpha_{S}^{-1}J_{K}^{-2}\rho^{-2}\right)\text{.}\end{equation}

The additional logarithmic term $\ln^{3}\left(\alpha_{S}^{-1}J_{K}^{-2}\rho^{-2}\right)$
occurs only near particle-hole symmetry and is replaced by a constant
of order unity away from particle hole-symmetry. The prefactor $\beta_{S}/\alpha_{S}$
grows as $S^{2}$ for large $S$. Thus, it is easily possible that
the quartic coefficient of the magnetic order parameter is significantly
enhanced in case of localized spins. For example, the relative corrections
are around $200\%$ for $S=7/2$ and $J_{K}\rho\simeq0.025$. Since
the order parameter coupling and the quartic coefficients of the superconducting
term are unchanged, it follows that the condition $\gamma<\sqrt{u_{s}u_{m}}$
for coexisting order can now be fulfilled easier than in the case
of purely itinerant systems. This offers a natural explanation for
the observation of homogeneous coexistence of both phases in systems
such as the $\mathrm{RNi_{2}B_{2}C}$ and addresses the fact that
coexistence observed in systems with localized spins is not in contradiction
to our conclusions. For completeness, we also analyzed a model with
additional magnetic interactions between localized spins that are
not captured by the RKKY mechanism, adding to $\mathcal{H}_{\mathrm{AFM}}$
of Eq.(\ref{H_AFM_local}) the term $\frac{1}{2}\sum_{i,j}J_{ij}\mathbf{S}_{i}\cdot\mathbf{S}_{j}$.
This new term will change the value of $T_{N}$, but not affect the
expression Eq.(\ref{um loc}) for $u_{m}$.

Additional consequences for localized spins are that the coefficient
$a_{m,0}=\left(\alpha_{S}J_{K}^{2}\right)^{-1}$ of the temperature
dependent quadratic coefficient $a_{m}=a_{m,0}\left(T-T_{N}\right)$
is expected to be larger compared to the corresponding coefficient
$a_{s,0}\simeq\rho/T_{c}$ of the superconducting order parameter
if we consider the multicritical point $T_{c}=T_{N}$ near particle-hole
symmetry, since $a_{m,0}/a_{s,0}\simeq\chi_{ph}\left(\mathbf{Q}\right)\ /\rho$.
With Eq.\ref{condition_suppression} follows then that it becomes
harder to achieve a suppression of the magnetization with $d\mathbf{M}^{2}/dT>0$
below the superconducting transition. The observed suppression\cite{Pratt09,Christianson09,Kreyssig10}
of $M$ in the coexistence region of $\mathrm{Ba\left(Fe_{1-x}Co_{x}\right)_{2}As_{2}}$
and $\mathrm{Ba\left(Fe_{1-x}Rh_{x}\right)_{2}As_{2}}$ is therefore
an indication that the same electrons are responsible for both states
and that magnetism is itinerant in these materials. On the other hand
the condition $a_{s,0}\gamma<a_{m,0}u_{m}$ for suppression of $T_{c}$
in the magnetically ordered state can easily be fulfilled. Thus, while
SC in systems with localized spins is affected by magnetic long range
order, the opposite does not seem to hold and AFM is rather indifferent
to SC.

Finally we comment on the relevance of this calculation for heavy
fermion superconductors, such as $\mathrm{CeRhIn_{5}}$ and $\mathrm{UPt_{3}}$. In $\mathrm{CeRhIn_{5}}$,
the coexistence between magnetism and superconductivity has
been investigated in great detail\cite{Knebel09}, while in $\mathrm{UPt_{3}}$ there 
is clear evidence\cite{Isaacs95} for suppresion of magnetism below $T_c$. The heavy fermion
system are believed to be properly described by the Kondo lattice
Hamiltonian\cite{Coleman07} with coupling between localized and
conduction electrons as in $\mathcal{H}_{\mathrm{AFM}}$ of Eq.(\ref{H_AFM_local}).
However, our analysis of this model, where we completely ignored the
Kondo effect and the emergence of a heavy electron state is inadequate
for such systems. In fact one expects that a system in the heavy electron
state is better described by the theory employed here for the FeAs
systems, yet the interactions $I$ and $V$ as well as the quasi-particles
masses are heavily renormalized due to Kondo lattice screening. Thus,
while a detailed theory for the competition of magnetism and superconductivity
in heavy electron states is complex, we do expect a similar competition
in the FeAs systems and these heavy fermion compounds. The very similar
behavior of the SC and AFM transition lines in the phase diagrams\cite{Fernandes10,Knebel09,Sachdev10_arxiv}
of $\mathrm{Ba\left(Fe_{1-x}Co_{x}\right)_{2}As_{2}}$ and $\mathrm{CeRhIn_{5}}$,
as well as the suppression of the magnetization\cite{Isaacs95} of
$\mathrm{UPt_{3}}$ below $T_{c}$, certainly support this view.

\section{Suppression of SC in the overdoped region}

So far we have analyzed the competition between AFM and SC with the
consequent suppression of the superconducting state in the underdoped
region of the phase diagram of the iron arsenides. In the overdoped
region there is no magnetically ordered state, yet SC is also suppressed
and eventually disappears. The Fermi surface in this part of the phase
diagram also changes significantly: for electron (hole) doped samples,
it is characterized by increasingly large electron (hole) pockets
and decreasingly small hole (electron) pockets\cite{Liu09}, which
eventually disappear at a certain doping level. In this section, we
investigate how the the disappearance of these pockets from the Fermi
surface affects the transition temperature of the pure SC state.

We consider, once more, one hole pocket centered at the Brillouin
zone and one electron pocket displaced by $\mathbf{Q}$ from the zone
center. For definiteness, we use the band dispersions (\ref{band_dispersion})
with $\varepsilon_{0}\equiv\varepsilon_{1,0}=\varepsilon_{2,0}$ and
vary the chemical potential $\mu$. Here, we consider the effects
of electron doping only, such that $\mu>0$; the case of hole doping
($\mu<0$) is analogous and the same conclusions hold. For simplicity,
we first consider $V$ to be constant as $\mu$ changes and neglect
intraband pairing interactions. The linearized gap equations are then
given by:

\begin{eqnarray}
\Delta_{1} & = & -\frac{V\Delta_{2}}{2}\int_{-W}^{W}d\xi\rho_{2}\left(\xi\right)\frac{\tanh\left(\frac{\xi}{2T_{c}}\right)}{\xi},\nonumber \\
\Delta_{2} & = & -\frac{V\Delta_{1}}{2}\int_{-W}^{W}d\xi\rho_{1}\left(\xi\right)\frac{\tanh\left(\frac{\xi}{2T_{c}}\right)}{\xi},\label{linearized_gap_SC}\end{eqnarray}
where, once again, $W$ denotes an upper energy cutoff associated
to the pairing interaction. In a two-dimensional system, the density
of states $\rho_{i}\left(\xi\right)$ is constant if the energy $\xi$
falls inside the bands. Therefore, diagonalization of the linearized
gap equations give the following implicit expression for $T_{c}$:

\begin{equation}
\frac{1}{\lambda_{0}^{2}}=\int_{-\frac{W}{2T_{c}}}^{\frac{\varepsilon_{0}-\mu}{2T_{c}}}dx\int_{-\min\left(\frac{W}{2T_{c}},\frac{\varepsilon_{0}+\mu}{2T_{c}}\right)}^{\frac{W}{2T_{c}}}dy\frac{\tanh x\:\tanh y}{4xy}\ \label{equation_Tc}\end{equation}
 where we introduced $\lambda_{0}=V\sqrt{\rho_{1}\rho_{2}}$. For
simplicity, let us first consider the special case $W=\varepsilon_{0}$;
the main conclusions hold for an arbitrary $W$.

Although a complete analytical solution for $T_{c}\left(\mu\right)$
is not available from Eq. (\ref{equation_Tc}), we can obtain some
important limits. For small $\mu\ll\varepsilon_{0}$, we obtain that
$T_{c}$ decreases linearly with respect to $T_{c}^{(0)}\equiv T_{c}\left(\mu=0\right)$:

\begin{equation}
T_{c}=T_{c}^{(0)}\left(1-\frac{\mu}{4\varepsilon_{0}}\right)\label{tc_small_mu}\end{equation}

Equivalently, we can show that the effective coupling constant $\lambda_{\mathrm{eff}}\equiv\ln^{-1}\left(\frac{W}{T_{c}}\frac{2\mathrm{e}^{\gamma_{E}}}{\pi}\right)$
decreases linearly with $\mu$ from its $\mu=0$ value $\lambda$.
At the special point $\mu=\varepsilon_{0}$, i.e. when the hole pocket
shrinks to a single point in the Fermi surface, the effective coupling
constant is reduced to about $70\%$ of its initial value, $\lambda_{\mathrm{eff}}=\lambda/\sqrt{2}$,
implying:

\begin{equation}
T_{c}=T_{c}^{(0)}\mathrm{e}^{-\frac{\sqrt{2}-1}{\lambda}}\label{tc_mu_epsilon0}\end{equation}

Even though the vanishing of the hole pocket does not cause $T_{c}$
to vanish, it does signal the onset of a regime where the coupling
constant decays very strongly with respect to the chemical potential
$\mu$, in contrast to the case of small chemical potential, where
the decay $\lambda_{\mathrm{eff}}$ was linear in $\mu$. To illustrate
this point, consider the regime of $\varepsilon_{0}<\mu<2\varepsilon_{0}$
such that $\mu-\varepsilon_{0}\gg T_{c}$. Notice that this condition
is not too restrictive, since $T_{c}<T_{c}^{(0)}\ll\varepsilon_{0}$,
by construction. Then, it follows that:

\begin{equation}
\lambda_{\mathrm{eff}}=\frac{\lambda^{2}}{2}\ln\left(\frac{\varepsilon_{0}}{\mu-\varepsilon_{0}}\right)\label{lambda_large_mu}\end{equation}
implying:

\begin{equation}
T_{c}=T_{c}^{(0)}\exp\left[-\frac{2}{\lambda^{2}\ln\left(\frac{\varepsilon_{0}}{\mu-\varepsilon_{0}}\right)}+\frac{1}{\lambda}\right]\label{tc_large_mu}\end{equation}

\begin{figure}

\begin{centering}
\includegraphics[width=0.8\columnwidth]{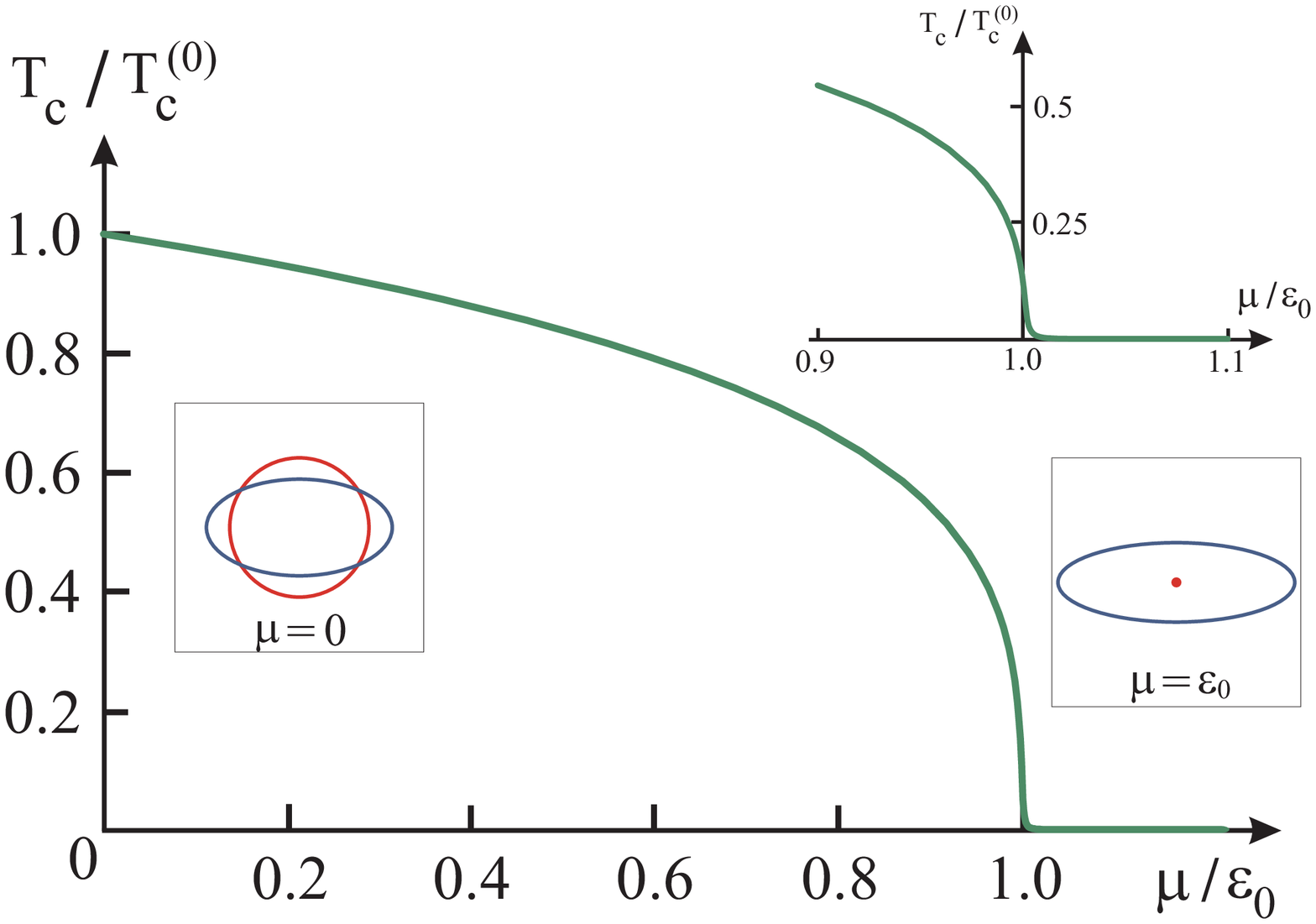} 
\par\end{centering}

\begin{centering}
\bigskip{}

\par\end{centering}

\begin{centering}
\includegraphics[width=0.8\columnwidth]{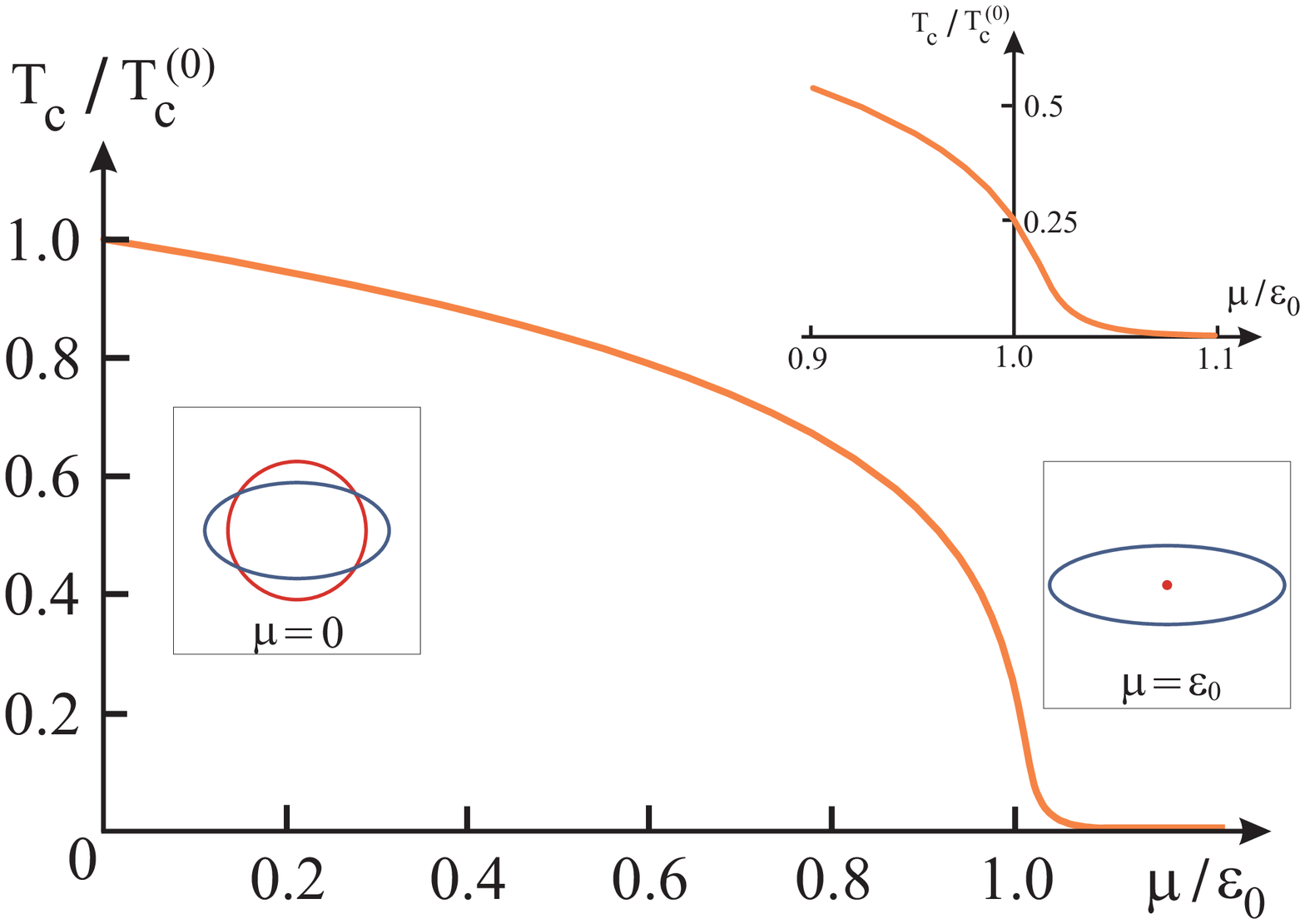} 
\par\end{centering}

\caption{Superconducting transition temperature $T_{c}$ (in units of its value
at zero chemical potential $T_{c}^{(0)}$) as function of the chemical
potential $\mu$ (in units of $\varepsilon_{0}$) for both an initial
effective SC coupling $\lambda_{0}=0.2$ (a) and $\lambda_{0}=0.3$
(b). The insets show the changes in the Fermi surface with $\mu$,
with the red circle (blue ellipse) denoting the hole (electron) pocket.
In the upper right corner of each panel we present a zoom of the region
around $\mu=\varepsilon_{0}$ where the electron pocket is reduced
to a Fermi point. }

\label{fig_tc_overdoped} 
\end{figure}

Clearly, the effective coupling constant only vanishes at $\mu=W+\varepsilon_{0}=2\varepsilon_{0}$,
i.e. where the range of energies for which the net attractive interaction
is positive does not cross the hole pocket. However, it is already
significantly reduced for values of the chemical potential much smaller
than that one. In figure \ref{fig_tc_overdoped}, we present the behavior
of $T_{c}$ for two values of the effective zero-doping coupling constant,
$\lambda_{0}=0.2$ and $\lambda_{0}=0.3$. Notice that, in both cases,
$T_{c}$ decreases moderately for $\mu<\varepsilon_{0}$ (i.e. when
the hole pocket is small but still present) and then is strongly reduced
for $\mu>\varepsilon_{0}$ (i.e. when the hole pocket disappears from
the Fermi surface). The same conclusions also hold for the case of
an arbitrary cutoff $W>\varepsilon_{0}$: the special point of the
phase diagram where the hole pocket is reduced to a Fermi point marks
the onset of a dramatic reduction of $T_{c}$, no matter the initial
value of the coupling constant $\lambda_{0}$.

As we mentioned, $T_{c}$ actually only vanishes at $\mu=W+\varepsilon_{0}$.
However, in this region where it is strongly suppressed, even an uniform
Coulomb interaction is able to completely destroy superconductivity.
This is to be contrasted to the optimally doped region, where the
$s^{+-}$ state is robust against an uniform Coulomb repulsion, as
we discussed in the previous section.

To illustrate how the $s^{+-}$ SC state is killed in the region of
strong suppression, $\mu-\varepsilon_{0}\gg T_{c}$, consider $W=\varepsilon_{0}$
and an uniform repulsion $U>0$. For simplicity, we focus only on the
limit of $U\gg V$. To leading order, the equation determining the
effective SC coupling constant is given by:

\begin{equation}
\frac{2\ln\left(\frac{\varepsilon_{0}}{\mu-\varepsilon_{0}}\right)V\rho_{1}\rho_{2}}{\lambda_{\mathrm{eff}}\ln\left(\frac{\varepsilon_{0}}{\mu-\varepsilon_{0}}\right)\rho_{1}+2\rho_{2}}=1\label{expansion_largest_eigenvalue}\end{equation}

Defining $\lambda_{i}=V\rho_{i}$, we obtain:

\begin{equation}
\lambda_{\mathrm{eff}}=\frac{2\lambda_{2}}{\lambda_{1}}\left[\lambda_{1}-\frac{1}{\ln\left(\frac{\varepsilon_{0}}{\mu-\varepsilon_{0}}\right)}\right]\label{coulomb_suppression}\end{equation}

Note that $\lambda_{\mathrm{eff}}$ vanishes at $\mu^{\ast}=\varepsilon_{0}\left(1+\mathrm{e}^{-1/\lambda_{1}}\right)$,
but is positive for $\mu<\mu^{\ast}$. Thus, a sufficiently large
Coulomb interaction $U$ is now able to destroy the $s^{+-}$ SC state,
contrary to what happened when the hole pocket was part of the Fermi
surface, where $\lambda_{\mathrm{eff}}\rightarrow\mathrm{const.}$
even as $U\rightarrow\infty$. 

\begin{figure}
\begin{centering}
\includegraphics[width=0.8\columnwidth]{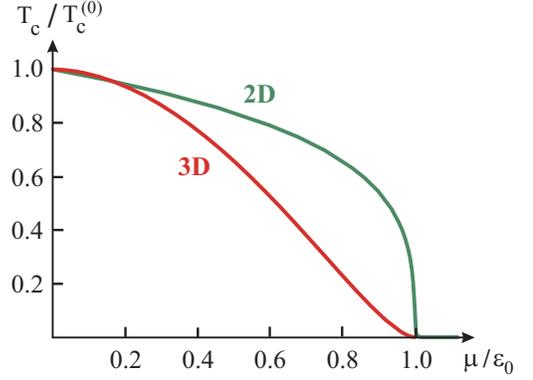}
\par\end{centering}

\caption{Comparison between the behavior of $T_{c}$ (in units of its value
at zero chemical potential $T_{c}^{(0)}$) as function of the chemical
potential $\mu$ (in units of $\varepsilon_{0}$) for a two-dimensional
and a three-dimensional system (green and red curves, respectively).
The coupling contants were chosen to give $T_{c}^{(0)}\approx0.03\: W$
in both cases, with $W=\varepsilon_{0}$. \label{fig_overdoped_3D}}

\end{figure}

The previous analysis holds for a $2D$ system. In the case of a three-dimensional
system, the density of states is not constant anymore, but given by
$\rho_{1}\left(\xi\right)=c_{1}\sqrt{\varepsilon_{0}-\mu-\xi}$ for
the hole band and by $\rho_{2}\left(\xi\right)=c_{2}\sqrt{\varepsilon_{0}+\mu+\xi}$
for the electron band. Here, we neglected the anisotropy of the electron
band and denoted unimportant constants by $c_{i}$ . The main difference
from the two-dimensional case is that the densities of states go to
zero at the bands edges, leading to a significant stronger reduction
of $T_{c}$ as the chemical potential increases. This is illustrated
in figure \ref{fig_overdoped_3D}, where we compare the solutions
of the linearized gap equations (\ref{linearized_gap_SC}) for the
cases of a $2D$ and a $3D$ system, with coupling constants chosen
to yield $T_{c}\left(\mu=0\right)\approx0.03\: W$. Thefore, close
to the point of the phase diagram where the hole pocket disappears,
the space dimensionality matters much more than in the usual Cooper
problem.

Our results suggest that the main factor responsible for the complete
suppression of the SC state in the overdoped region of the iron arsenides
phase diagram is the evolution of the Fermi surface with doping. This
is corroborated by ARPES measurements\cite{Sekiba09,Brouet09,Liu10Lif}
on $\mathrm{Ba\left(Fe_{1-x}Co_{x}\right)_{2}As_{2}}$, which show
no superconductivity in the overdoped region after the Fermi surface
loses one of its pockets around $x\approx15\%$. The fact that superconductivity
requires the presence of the hole pocket and the observation that
the superconducting gap on the hole and electron pockets are very
similar, strongly support the view that the pairing interaction in
the pnictides is due to interband coupling.

\section{Concluding remarks}

We showed that the ability of superconductivity and antiferromagnetism
to order simultaneously depends sensitively on the nature of the Cooper
pair wave function. In a two-band system with particle-hole symmetry,
the itinerant AFM state and the unconventional $s^{+-}$ SC state
are exactly in a borderline regime between phase coexistence and phase
separation. In contrast, the conventional $s^{++}$ state is deep
in the regime of mutual phase exclusion. We further demonstrated that
this result holds regardless of additional details of the band structure
or the system's dimensionality. It does not change either when one
considers the presence of intraband pairing interactions or the effects
of Coulomb repulsion. The robustness of this result, valid around the multicritical point $T_N \simeq T_c$, is related to
the quasiparticle excitation spectrum of the system, which depends
on the peculiar combination $\mathbf{M}^{2}+\left\vert \Delta\right\vert ^{2}\equiv\overrightarrow{\mathbf{N}}^{2}$
only. This is the root of the SO$(5)$ symmetry of the free energy
expansion, which has been shown to hold not only in the mean-field
level, but also in the strong coupling limit, as a subgroup of an
emergent SO$(6)$ symmetry present in the Hamiltonian\cite{Podolski09}.
Furthermore, the inclusion of fluctuations in our free energy expansion
are known to not change the condition for having the borderline regime,\cite{Kosterlitz76,Aharony03}
$g=0$. All these facts suggest that this simple result is a much
more general property of this type of system. We also demonstrated
that for our results to hold, it is crucial that the same electrons
that form Cooper pairs are responsible for the formation of the ordered
moment. For instance, we showed that an AFM state generated by localized
moments has a tendency of being indifferent to SC, falling much easier
in the regime of phase coexistence.

When applying this result to the pnictides, one has to critically
evaluate whether the assumptions that were made are too restrictive.
We assumed a certain vicinity to particle-hole symmetry. It is evident
that the real materials do not have perfectly nested bands\cite{Liu09}.
Yet, there are clearly two sets of pockets whose centers are separated
by the ordering vector $\mathbf{Q}$. As we showed, small perturbations
in the ellipticity of one pocket or in the chemical potential bring
the $s^{+-}$ state to the regime of mutual exclusion. However, simultaneous
perturbations in both quantities can bring the system to either regime,
as our figure \ref{fig_gpp_gpm} illustrates. Most importantly, small
deviations from perfect nesting are not sufficient to lead to simultaneous
order of magnetism and $s^{++}$-pairing close to the multicritical point. The fact that particular
details of the band structure are able to bring the $s^{+-}$ state
either to the phase coexistence or to the phase separation regime
is, in our view, what makes some of the iron arsenides display second-order
AFM-SC transition and others, first-order AFM-SC transition. For $\mathrm{Ba\left(Fe_{1-x}Co_{x}\right)_{2}As_{2}}$,
we were able to independently {}``fit'' the effective band structure
parameters to give the correct doping and $T_{N}$ dependence of the
zero-temperature magnetization in the absence of superconductivity\cite{Fernandes10}.
Having this set of parameters, which did not depend on any SC property,
we were able to obtain the complete mean-field $x-T$ phase diagram and to readily calculate the parameter $g$, verifying that,
as evidenced by many experimental probes,\cite{Ni09,Chu09,Laplace09,Julien09,Bernhard09,Ning09,Lester09,Pratt09,Christianson09}
$\mathrm{Ba\left(Fe_{1-x}Co_{x}\right)_{2}As_{2}}$ has homogeneous
coexistence between SC and AFM states.

For other compounds that are believed to display phase separation,
like $\mathrm{LaFeAs\left(O_{1-x}F_{x}\right)}$ and $\mathrm{\left(Ba_{1-x}K_{x}\right)Fe_{2}As_{2}}$,
we do not have the same systematic diffraction measurements that allow
a reliable extraction of effective two-band parameters. Even though
tight-binding fits to DFT-calculated band structures are available,
they usually refer to the parent compounds. The problem to extrapolate
them to finite doping is that, in these materials, doping is not on
the Fe site nor on the FeAs plane, which makes it much more difficult
to make a direct association between electronic occupation number
and doping level, as we did for $\mathrm{Ba\left(Fe_{1-x}Co_{x}\right)_{2}As_{2}}$.
Furthermore, in some of the pnictides, the issue of whether they have
homogeneous or heterogeneous AFM-SC coexistence is still not completely
settled. For example, in $\mathrm{\left(Ba_{1-x}K_{x}\right)Fe_{2}As_{2}}$,
while many experiments find phase separation, M\"ossbauer spectroscopy
identifies microscopic coexistence\cite{Rotter09}. In $\mathrm{SmFeAs\left(O_{1-x}F_{x}\right)},$
muon spin rotation\cite{Drew09} finds homogeneous coexistence while
a work combining powder x-ray diffraction, M\"ossbauer spectroscopy
and nuclear resonant forward scattering finds phase separation\cite{Kamihara10}.

In what concerns the pairing state, although many theories for pairing
due to electronic interactions predict a nodeless $s^{+-}$ state,
other models also suggest that accidental (i.e. not related to symmetry)
nodes\cite{Sknepnek09,Chubukov09_nodes,Thomale09} or even d-wave
nodal states\cite{Graser09} could be present. Furthermore, penetration depth experiments are in principle consistent with
the existence of accidental nodes \cite{Gordon09}. The inclusion of gap
nodes in our model, as we discussed, tend to move the SC state from
the borderline regime to the mutual exclusion regime, although not
as deep as in the $s^{++}$ case. Clearly, $A_{1g}$ sign-changing
gap functions with specific configurations of nodes could be as effective
as small perturbations of the particle-hole symmetric band structure
in making the AFM-SC state go from one regime to the other. Thus,
while we cannot discard the presence of unconventional nodal states
coexisting with AFM, we can certainly discard the conventional $s^{++}$
state: it is simply incompatible with magnetism.

Regarding the nature of the magnetic state, there is still some debate
whether the AFM phase is due to conduction electrons or localized
spins. Our analysis shed light on this subject: as our calculations
demonstrated, a magnetically ordered phase with localized spins can
coexist much easier with superconductivity. This holds even for conventional
BCS states, like in the case of the ternary\cite{Maple82} and quaternary\cite{Canfield98}
rare earth compounds. Furthermore, the magnetization is not so affected
by the SC condensate as in the case of purely itinerant magnetism.
Therefore, the experimental observation of reentrance of the paramagnetic
phase inside the SC dome\cite{Fernandes10} seems to rule out the
AFM state formed only by localized moments.

As we showed, in the underdoped side of the FeAs phase diagram, both
$T_{N}$ and $T_{c}$ are suppressed due to the competition between
the AFM and SC phases. The N\'eel transition line is even bent back,
approaching the $x$ axis vertically, similar to the case of a first-order
transition line. The suppression of $T_{c}$ is milder; this difference
is probably due to the fact that the AFM gap opens only around the
corresponding Bragg scattered points,\cite{Parker_Vavilov09} whereas
the SC gap opens isotropically around the Fermi surface. In the overdoped
side, however, SC is the only thermodynamic ordered phase of the system
and yet it is suppressed similarly to the underdoped side. Our calculations
using the coupled SC gap equations shows that this suppression is
related to the disappearance of one of the pockets from the Fermi
surface. As suggested by our figures \ref{fig_tc_overdoped} and \ref{fig_overdoped_3D}, and by
ARPES measurements\cite{Sekiba09,Brouet09,Liu10Lif} in $\mathrm{Ba\left(Fe_{1-x}Co_{x}\right)_{2}As_{2}}$,
the main cause of suppression of SC in the overdoped side seems to
be the doping-induced changes in the Fermi surface rather than possible
changes in the magnitude of the pairing interaction. Interestingly,
the Fermi surface is also indirectly responsible for the suppression
of $T_{c}$ in the underdoped side, since it is the driving force
of the AFM instability\cite{Eremin10}.

The calculations we presented here did not take into account another
important degree of freedom present in the iron arsenides, the orthorhombic
distortion. However, in separate works, we showed that the competition
between SC and AFM, combined with the coupling between nematic degrees
of freedom and structural distortion, are able to consistently explain
the observed back-bending of the structural transition line\cite{Nandi10}
inside the SC dome as well as the increase of the shear modulus\cite{Fernandes09}
below $T_{c}$.

Finally, we comment that in many models of unconventional pairing
in the iron arsenides, the bosons responsible for the formation of
Cooper pairs are spin fluctuations\cite{Mazin08,Kuroki08,Chubukov08,Sknepnek09,Graser09,Junhua09}.
In this case, the pairing interaction $V$ itself would also be sensitive
to the presence of magnetic long range order. Clearly, this would
change some details of the coexistence state, particularly the form
of the transition lines. Yet, our main conclusions still hold in this
case, since the decision about the coexistence between AFM and SC
is made when both transition temperatures are close, implying that
the order parameters are only infinitesimal.

\bigskip{}

We thank S. Bud'ko, P. Canfield, P. Chandra, A. Chubukov, A. Goldman,
D. Johnston, A. Kaminski, A. Kreyssig, R. McQueeney, D. Pratt, R.
Prozorov, S. Sachdev and M. Vavilov for fruitful discussions. This
work was supported by the U.S. DOE, Office of BES, DMSE. Ames Laboratory
is operated for the U.S. DOE by Iowa State University under Contract
No. DE-AC02-07CH11358.

\appendix

\section{Diagrammatic interpretation of the Ginzburg-Landau coefficients}

Here, we rederive the Ginzburg-Landau expansion in Eq. (\ref{F})
by explicitly integrating out the fermionic degrees of freedom of
a system with competing SC and AFM. This method is useful since it
provides a diagrammatic interpretation for the coefficients, as presented
in figure \ref{fig_feynman}. We first define the Nambu operator:

\begin{equation}
\psi_{\mathbf{k}}=\left(\begin{array}{cccc}
c_{\mathbf{k}\uparrow} & c_{-\mathbf{k}\downarrow}^{\dagger} & d_{\mathbf{k+Q}\uparrow} & d_{-\mathbf{k}-\mathbf{Q}\downarrow}^{\dagger}\end{array}\right)^{T}\label{Nambu_operator}\end{equation}
 and generalize the uniform order parameters $\Delta_{\alpha}$ and
$M$ to inhomogeneous functions of space and time $\Delta_{\alpha,\left(\mathbf{k},\omega_{n}\right)}$
and $M_{\left(\mathbf{k},\omega_{n}\right)}$, where $\omega_{n}=(2n+1)\pi T$
is a fermionic Matsubara frequency. Thus, denoting $k=\left(\mathbf{k},\omega_{n}\right)$,
we obtain the Green's function:

\begin{equation}
\hat{\mathcal{G}}_{k,k^{\prime}}^{-1}=\left(i\omega_{n}-\hat{\xi}_{\mathbf{k}}\right)\delta_{k,k^{\prime}}-\hat{U}_{k-k^{\prime}}\label{fermionic_Green_function}\end{equation}
 where the hat denotes a matrix in Nambu space and:

\begin{equation}
\hat{U}_{k-k^{\prime}}=\hat{\Delta}_{k-k^{\prime}}+\hat{M}_{k-k^{\prime}}\label{aux_fermionic_Greens_function}\end{equation}

With the help of the Pauli matrices $\boldsymbol{\tau}$, we can write
the $4\times4$ Nambu matrices as

\begin{equation}
\hat{\xi}_{\mathbf{k}}=\left(\begin{array}{cc}
\xi_{1,\mathbf{k}}\tau_{z} & 0\\
0 & \xi_{2,\mathbf{k}+\mathbf{Q}}\tau_{z}\end{array}\right)\label{nambu_matrix_xi}\end{equation}
 and

\begin{equation}
\hat{\Delta}_{q}=\left(\begin{array}{cc}
-\Delta_{1,q}\tau_{x} & 0\\
0 & -\Delta_{2,q}\tau_{x}\end{array}\right)\label{nambu_matrix_delta}\end{equation}
 as well as

\begin{equation}
\hat{M}_{q}=\left(\begin{array}{cc}
0 & -M_{q}\tau_{0}\\
-M_{q}\tau_{0} & 0\end{array}\right)\label{nambu_matrix_M}\end{equation}

Therefore, after introducing the condensation energy of the magnetic
and superconducting phases, we obtain the action:

\begin{eqnarray}
S & = & \int_{k}\psi_{k}^{\dagger}\hat{\mathcal{G}}_{k,k^{\prime}}^{-1}\psi_{k^{\prime}}^{\dagger}\nonumber \\
 &  & +2\int_{x}\left[\frac{M^{2}}{I}-\frac{\Delta_{1}\Delta_{2}}{V}\right]\label{total_action}\end{eqnarray}
 where $\int_{k}=T\sum\limits _{n}\sum\limits _{\mathbf{k}}$ and
$\int_{x}=\frac{1}{v}\int d^{d}x\int_{0}^{T^{-1}}d\tau$, with $v$
denoting the volume of the system. The fermions can now be integrated
out, yielding an effective action in terms of the collective AFM and
SC fields $S_{\mathrm{eff}}$. The partition function of the free
fermions is given by:

\begin{equation}
Z_{0}=\mathrm{det}\left(-\hat{\mathcal{G}}_{0}^{-1}\right)\label{Z_free_fermions}\end{equation}
 where we defined the non-interacting Nambu Green's function $\hat{\mathcal{G}}_{0,kk^{\prime}}^{-1}=\left(i\omega_{n}-\hat{\xi}_{\mathbf{k}}\right)\delta_{k,k^{\prime}}$.
The effective action reads:

\begin{equation}
S_{\mathrm{eff}}=-\mathrm{Tr}\ln\left(1-\hat{\mathcal{G}}_{0}\hat{U}\right)+2\int_{x}\left(\frac{M}{I}-\frac{\Delta_{1}\Delta_{2}}{V}\right)\label{effective_action}\end{equation}

Here, the trace refers to the sum over momentum, frequency and Nambu
indices. Notice that, for uniform AFM and SC gaps, the total free
energy density (free-fermions contribution included) is given by:

\begin{equation}
f=-\int_{k}\ln\:\det\left(\hat{\mathcal{G}}_{k,k^{\prime}}^{-1}\right)+\frac{2M^{2}}{I}-\frac{2\Delta_{1}\Delta_{2}}{V}\label{aux_free_energy}\end{equation}

Since $\det\left(\hat{\mathcal{G}}_{k,k^{\prime}}^{-1}\right)=\sum\limits _{\mathbf{k},a}\left(\omega_{n}^{2}+E_{a,\mathbf{k}}^{2}\right)$,
we can evaluate the Matsubara sum to obtain:

\begin{equation}
f=-2T\sum\limits _{\mathbf{k},a}\ln\left[2\cosh\left(\frac{E_{a,\mathbf{k}}}{2T}\right)\right]+\frac{2M^{2}}{I}-\frac{2\Delta_{1}\Delta_{2}}{V}\label{free_energy}\end{equation}
 with excitation energies $E_{a,\mathbf{k}}$ given by the positive
roots of (\ref{excitation_energies}). Minimization of the free energy
(\ref{free_energy}) with respect to $\Delta_{\alpha}$ and $M$ leads
then to the gap equations (\ref{gap_equations}).

Going back to the effective action (\ref{effective_action}), we can
now perform a Ginzburg-Landau expansion in the AFM and SC order parameters.
Expansion of the logarithm yields, for the free energy relative to
the paramagnetic, normal phase:

\begin{eqnarray}
\delta F & = & \frac{1}{2}\mathrm{Tr}\left(\hat{\mathcal{G}}_{0}\hat{U}\right)^{2}+\frac{1}{4}\mathrm{Tr}\left(\hat{\mathcal{G}}_{0}\hat{U}\right)^{4}\nonumber \\
 &  & +2\int_{x}\left(\frac{M^{2}}{I}-\frac{\Delta_{1}\Delta_{2}}{V}\right)\label{aux_free_energy_expansion}\end{eqnarray}

Considering static and homogeneous AFM and SC gaps and performing
the traces in the Nambu space, we obtain the free energy density expansion
of Eq. (\ref{F}) with Ginzburg-Landau coefficients:

\begin{eqnarray}
a_{m} & = & \frac{4}{I}+4\int_{k}G_{1,k}G_{2,k}\nonumber \\
a_{s,\alpha\beta} & = & -\frac{2}{V}\left(1-\delta_{\alpha\beta}\right)-2\delta_{\alpha\beta}\int_{k}G_{\alpha,k}G_{\alpha,-k}\nonumber \\
u_{m} & = & 4\int_{k}G_{1,k}^{2}G_{2,k}^{2}\nonumber \\
u_{s,\alpha} & = & 2\int_{k}G_{\alpha,k}^{2}G_{\alpha,-k}^{2}\nonumber \\
\gamma_{\alpha\alpha} & = & -4\int_{k}G_{\alpha,k}G_{\alpha,-k}G_{1,k}G_{2,k}\nonumber \\
\gamma_{\alpha\bar{\alpha}} & = & 2\int_{k}G_{\alpha,k}G_{\alpha,-k}G_{\bar{\alpha},k}G_{\bar{\alpha},-k}\label{formal_coefficients}\end{eqnarray}

In the previous expressions, we introduced the non-interacting single-particle
Green's functions $G_{1,k}=\left(i\omega_{n}-\xi_{1,\mathbf{k}}\right)^{-1}$
and $G_{2,k}=\left(i\omega_{n}-\xi_{2,\mathbf{k+Q}}\right)^{-1}$.
Thus, the coefficients depend only on the band structure and on the
magnitude of the electronic interactions. The Feynman diagrams associated
to the quartic coefficients of (\ref{formal_coefficients}) are presented
in figure \ref{fig_feynman}. Evaluation of the Matsubara sums then
leads to (\ref{GL_coefficients_1}), (\ref{GL_coefficients_2}) and
(\ref{GL_coefficients_3}) for the Ginzburg-Landau coefficients.


\begin{thebibliography}{10}
\bibitem{comment01} Notice that this quantity is a factor of $2$
greater than the $g$ we defined in our previous work, Ref. 20.
\end{thebibliography}
\end{document}